\documentclass[twocolumn]{aastex631}

\shorttitle{Starspot Evolution of $\zeta$~And}
\shortauthors{Roettenbacher et al.}

\begin{document}

\title{Interferometric Images of the Starspot Evolution  of $\zeta$~Andromedae }

\newcommand{\michigan}{Department of Astronomy, University of Michigan, Ann Arbor, MI 48109, USA}
\newcommand{\heidelberg}{Max-Planck-Institut f\"ur Astronomie (MPIA), K\"onigstuhl 17, 69117 Heidelberg, Germany}
\newcommand{\exeter}{Astrophysics Group, Department of Physics \& Astronomy, University of Exeter, Stocker Road, Exeter EX4 4QL, UK}
\newcommand{\grenoble}{Institut de Planetologie et d’Astrophysique de Grenoble UGA/CNRS, Grenoble F-38058, France}
\newcommand{\chara}{The CHARA Array of Georgia State University, Mount Wilson Observatory, Mount Wilson, CA 91203, USA}
\newcommand{\tsu}{Tennessee State University (retired), Nashville, TN  37209  USA}

\newcommand{\cires}{Cooperative Institute for Research in Environmental Sciences at the University of Colorado Boulder, Boulder, CO, 80309 USA}

\newcommand{\konkoly}{Konkoly Observatory, HUN-REN Research Centre for Astronomy and Earth Sciences, MTA Centre of Excellence, Konkoly Thege Miklós út 15-17., H-1121 Budapest, Hungary}

\newcommand{\szeged}{Department of Experimental Physics, Institute of Physics, University of Szeged, D\'om t\'er 9, Szeged, 6720 Hungary}

\newcommand{\lehigh}{Department of Physics, Lehigh University, Bethlehem, PA, 18015 USA}
\newcommand{\nasahq}{Astrophysics Division, NASA Headquarters, Washington, DC 20546, USA}

\newcommand{\aavso}{American Association of Variable Star Observers, USA}

\newcommand{\milkyway}{MTA-ELTE Lend{\"u}let ``Momentum'' Milky Way Research Group, Hungary}

\correspondingauthor{R.\ M.\ Roettenbacher}
\email{rmroett@umich.edu}

\author[0000-0002-9288-3482]{Rachael M.\ Roettenbacher}
\affiliation{\michigan}

\author[0000-0002-3380-3307]{John D.\ Monnier}
\affiliation{\michigan}

\author[0000-0003-0529-1161]{Heidi Korhonen}
\affiliation{\heidelberg}

\author[0000-0003-4155-8513]{Gregory W. Henry} 
\affiliation{\tsu}

\author[0000-0002-8351-6541]{Cliff Kotnik}
\affiliation{\aavso}

\author[0000-0002-3827-8417]{Joshua Pepper}
\affiliation{\lehigh}
\affiliation{\nasahq}

\author[0000-0002-3658-2175]{B\'alint Seli}
\affiliation{\konkoly}

\author[0000-0002-6471-8607]{Kriszti\'an Vida}
\affiliation{\konkoly}

\author[0000-0002-8585-4544]{Attila B\'odi}
\affiliation{Department of Astrophysical Sciences, Princeton University, 4 Ivy Lane, Princeton, NJ 08544, USA}

\author[0000-0002-6497-8863]{Borb\'ala Cseh}
\affiliation{\konkoly}
\affiliation{\milkyway}

\author[0000-0001-9210-9860]{G\'eza Cs\"ornyei}
\affiliation{European Southern Observatory, Karl-Schwarzschild-Straße 2, 85748 Garching, Germany}




\author[0000-0002-8813-4884]{M\'at\'e Krezinger}
\affiliation{\konkoly}

\author[0000-0002-8770-6764]{R\'eka K\"onyves-T\'oth}
\affiliation{\konkoly}
\affiliation{\szeged}

\author[0000-0002-1792-546X]{Levente Kriskovics}
\affiliation{\konkoly}

\author[0000-0003-0926-3950]{Kriszti\'an S\'arneczky}
\affiliation{\konkoly}

\author[0000-0001-7806-2883]{\'Ad\'am S\'odor}
\affiliation{\konkoly}

\author[0000-0002-1698-605X]{R\'obert Szak\'ats}
\affiliation{\konkoly}


\author[0000-0001-6017-8773]{Stefan Kraus}
\affiliation{\exeter}


\author[0000-0002-2208-6541]{Narsireddy Anugu}
\affiliation{\chara}

\author[0000-0001-9764-2357]{Claire L.\ Davies}
\affiliation{\exeter}

\author[0000-0002-3003-3183]{Tyler Gardner}
\affiliation{\cires}

\author[0000-0001-9745-5834]{Cyprien Lanthermann}
\affiliation{\chara}

\author[0000-0001-5415-9189]{Gail H.\ Schaefer}
\affiliation{\chara}

\author[0000-0001-5980-0246]{Benjamin R.\ Setterholm}
\affiliation{\heidelberg}


\begin{abstract}

The evolution of starspots of the giant primaries of RS CVn systems is typically detected indirectly with photometric and spectroscopic monitoring.  These observations suggest slowly-evolving stellar surfaces and can constrain differential rotation as starspots move with respect to one another. However, starspot latitudes are difficult to constrain without resolved images of the stellar surfaces from which the unambiguous locations of starspots  are determined. 
We imaged the active RS CVn primary $\zeta$~And with the 330-m-baseline Center for High Angular Resolution Astronomy Array for three epochs over approximately six rotations of the star.  The resultant images show a more complicated picture of stellar activity than expected from the contemporaneous photometry and earlier Doppler images.  The spot structures change on the timescale of rotation, making differential rotation difficult to study.  
Our observations show changes in the polar spot, growing over time.  We do not detect the secondary star in the interferometric data, though the observations are sensitive to the predicted 0.75~$M_\odot$ main-sequence star, and we  suggest the companion may be a white dwarf. 

\end{abstract}

\keywords{long baseline interferometry (932), optical interferometry (1168), magnetic variable  stars (996), close binary stars (254), starspots  (1572)}

\section{Introduction} \label{sec:intro}

For decades, ground-based photometric observations have been obtained for stars with rotational modulations.  These modulations are attributed to dark regions, or starspots, moving in and out of view as a star rotates \citep[e.g.,][]{kron47,kron50}.  The starspots are  believed to be the result of strong magnetic fields suppressing convection in the outer layers of the stars, analogous to sunspots.  Among the stars exhibiting starspots, the largest spots are often detected on RS Canum Venaticorum stars \citep[RS CVns;][]{strassmeier1999}.  As defined by \citet{hall1976}, these systems typically consist of an evolved primary star, usually a giant, and a less-evolved secondary star.  The stellar rotation of the primary and orbital period are often tidally locked.  The tidal forces with the companion spin up the primary star giving it more rapid rotation than for a single star of its age.  
The size of the primary,  proximity of the secondary, and  mass of both components in these systems constrain the distortion of the primary.  The primary stars  partially fill their Roche lobes, making the stars ellipsoidal.    

To better understand the  magnetic activity, stellar surfaces are reconstructed or imaged.  
Light-curve inversion is a method that reconstructs a stellar surface based upon one or more photometric light curves and provided known stellar parameters, including photosphere and spot temperatures, stellar inclination, and limb-darkening coefficients \citep[see, e.g.,][]{harmon00,vida10,roettenbacher11}.  
While the method can be used to determine starspot longitudes, the absolute latitude of starspots cannot be determined.  Some relative latitude information can be extracted, especially if contemporaneous light curves in multiple bandpasses are inverted together taking advantage of limb-darkening information.  Light-curve inversion is an ill-posed problem, with a family of degenerate solutions capable of reproducing the observed light curves.  To reduce the degeneracy, input stellar parameters, regularization, and other algorithm-specific choices are made for each data set to lead to a single reconstructed surface.

Spectroscopic Doppler imaging reconstructs a stellar surface from a time series of high-resolution stellar spectra \citep[see, e.g.,][]{vogt87,rice89}.  The method locates starspots on the stellar surface through tracing the motion of cool features in temperature-sensitive absorption lines as the star rotates.  Stellar parameters like photosphere and spot temperatures are determined from the data.  Some latitude information can be extracted from where absorption lines are impacted (e.g., wings versus the core). 
However, there is still the possibility of degeneracies between hemispheres of the star.  
Like light-curve inversion, Doppler imaging is an ill-posed problem that also requires additional information to break degeneracies to result in a single reconstructed surface. 

Directly imaging a stellar surface is possible with 
long-baseline optical interferometry.  The method uses widely-separated telescopes to effectively make a telescope with an equivalent diameter equal to the largest distance between pairs.  Sufficiently large interferometers are capable of not only resolving the star itself, but also features on its surface.  With aperture synthesis imaging of the observations, stars can be viewed as they appear on the sky \citep[see, e.g.,][]{monnier07,roettenbacher16b}, eliminating the degeneracies of the other methods, including the ambiguity of the latitudes of starspots.  
Analogous to the other methods, observations from an entire stellar rotation may be combined to obtain an image of the stellar surface.  This technique further provides parameters like the inclination and position angles of the rotation axis.
\citet{roettenbacher17b} performed a comparison of these three methods, showing the techniques have good agreement for the longitudes of starspots, but there were disagreements across the methods in latitudes.  Those results were consistent with our understanding of limitations in measuring latitudes.  

The RS CVn binary $\zeta$~Andromedae ($\zeta$~And; HD 4502) has long been a target for starspot studies investigating what has appeared to be slow evolution of the stellar surface's large starspots \citep[e.g.,][]{korhonen10,kovari2012}.  $\zeta$~And is a bright single-lined spectroscopic binary, first noted by Adelade Hobe and reported in \citet{campbell1911}  \citep[$H = 1.62$, $V = 4.06$;][]{ducati02,cutri03}. The system's stars are in a circular orbit of $P_\mathrm{orb} = 17.769426 \pm 0.000040$ days \citep{fekel1999} and are tidally locked, so the orbital period is taken to be the primary star's rotation period.  The K1 giant primary is known to be ellipsoidally variable \citep{stebbins1928} with the ratio of the equatorial axis to the polar axis of 1.06 \citep{roettenbacher16b}.

From stellar evolution grids, \citet{kovari07} determined the primary to be a red-giant branch star with a mass of $2.6 \pm 0.4 \ M_\odot$.  They estimated that the main-sequence secondary has a mass of $\approx 0.75 \ M_\odot$,  based on the mass ratio, $q = 0.29$, given by \citet{stawikowski1994}.  Consistent with these values, \citet{fekel1999} independently give a mass function of $f(m) = 0.02923 \pm 0.00012 \ M_\odot$.  The secondary star has not been spectroscopically detected.  

Repeated Doppler images of $\zeta$~And from data spanning 49 days (under three rotation periods) led to a measurement of the surface shear parameter (dimensionless) $\alpha = 0.055$ where $\alpha = (\Omega_\mathrm{eq} - \Omega_\mathrm{pole})/\Omega_\mathrm{eq}$ and $\Omega$ represents the rate of rotation in degrees per day \citep{kovari2012}.  This parameter characterizes differential rotation, which describes how the stellar rotation varies at different latitudes. $\zeta$~And was first interferometrically imaged by \citet{roettenbacher16b}.  In two epochs separated by about two years, the surface was observed to change significantly with transient starspots at low- and mid-latitudes.  Unlike the Sun, $\zeta$~And hosted a polar spot that had also evolved between the epochs.  The work revealed that $\zeta$~And likely has a dynamo different from that of the Sun in order to create the polar and asymmetric surface structures not reflected in the Sun's active low latitudes.  

In this article, we produce and analyze a new series of interferometric images of the spotted primary star of $\zeta$~And.  We compare our observations to past images, discuss the difficulties of measuring differential rotation, and investigate the nature of the secondary.  We describe the observations used in this study in Section 2.  In Section 3, we detail our images and our analysis.  The nature of the companion is described in Section 4.   We discuss our conclusions in Section 5.

\section{Observations} \label{sec:obs}

\subsection{MIRC-X Observations}
\label{section:mircx}

Interferometric observations were obtained at the  Center for High Angular Resolution Astronomy (CHARA) Array, which has a maximum baseline of 330~m and is located at Mount Wilson Observatory, CA \citep{tenbrummelaar05}.  The $H$-band Michigan InfraRed Combiner-eXeter \citep[MIRC-X;][]{anugu20} was used with all 6 CHARA Array telescopes.  The observations were obtained in grism mode ($R \sim 190$).  The combination of the CHARA Array and MIRC-X provides an angular resolution of $\sim 0.5$ mas.

\begin{figure}
    \includegraphics[trim={4.cm 2.5cm 2.5cm 2.5cm},clip,width=0.475\textwidth]{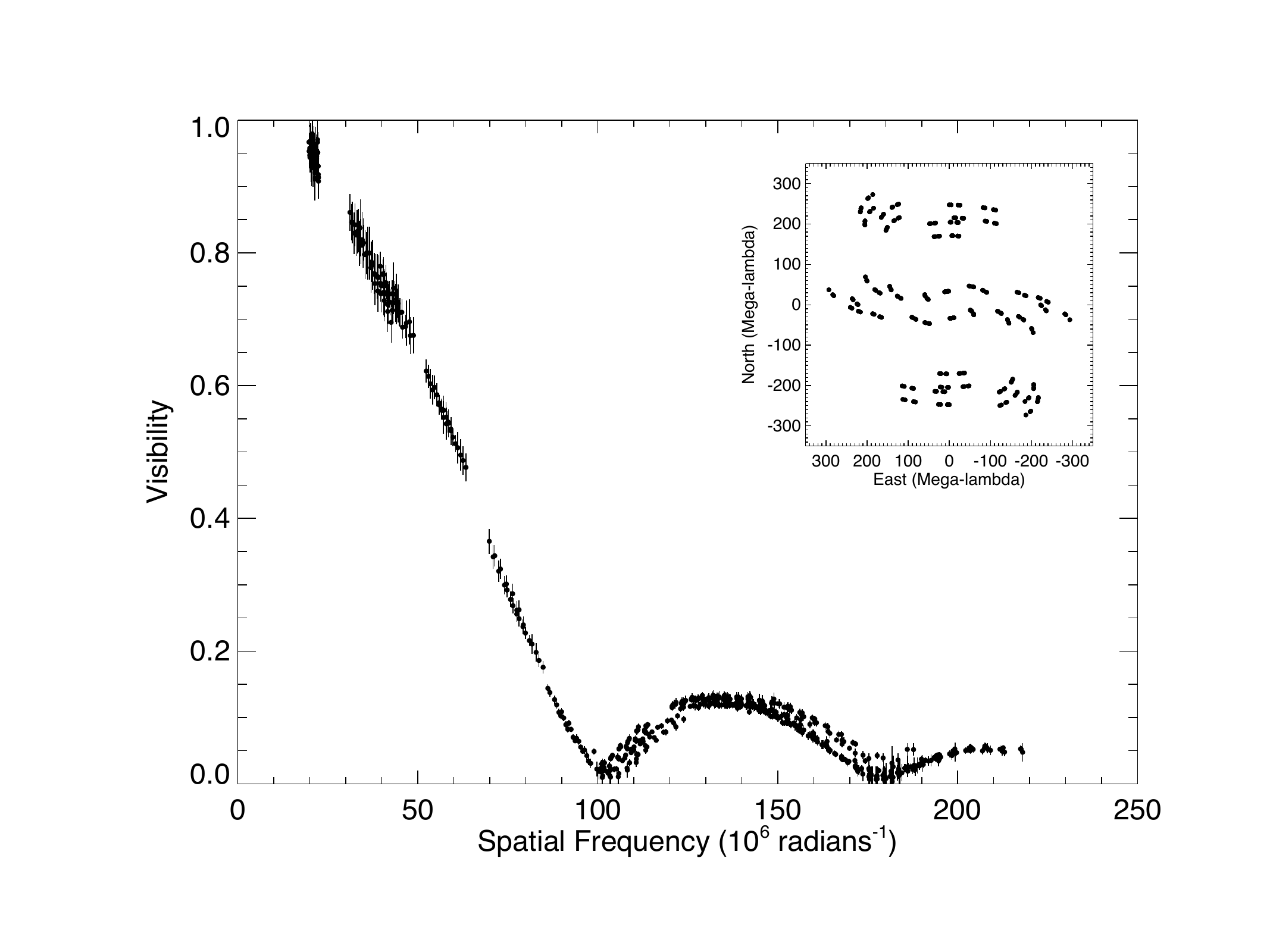}
    \caption{Visibility curve of UT 2019 July 17 $\zeta$~And CHARA/MIRC-X observations with $1\sigma$ error bars.  Inset is the $uv$ coverage for the night of observation.  The ellipsoidal shape of the star (data shown are at phase 0.98), limb darkening, and starspots contribute to the spread in visibility at a particular spatial frequency. }
     
    \label{fig:viscurve}
\end{figure}

\begin{figure*}
\begin{center}
    \includegraphics[angle=270,trim={1.5cm 1.5cm 1.5cm 1.5cm},clip,width=0.9\textwidth]{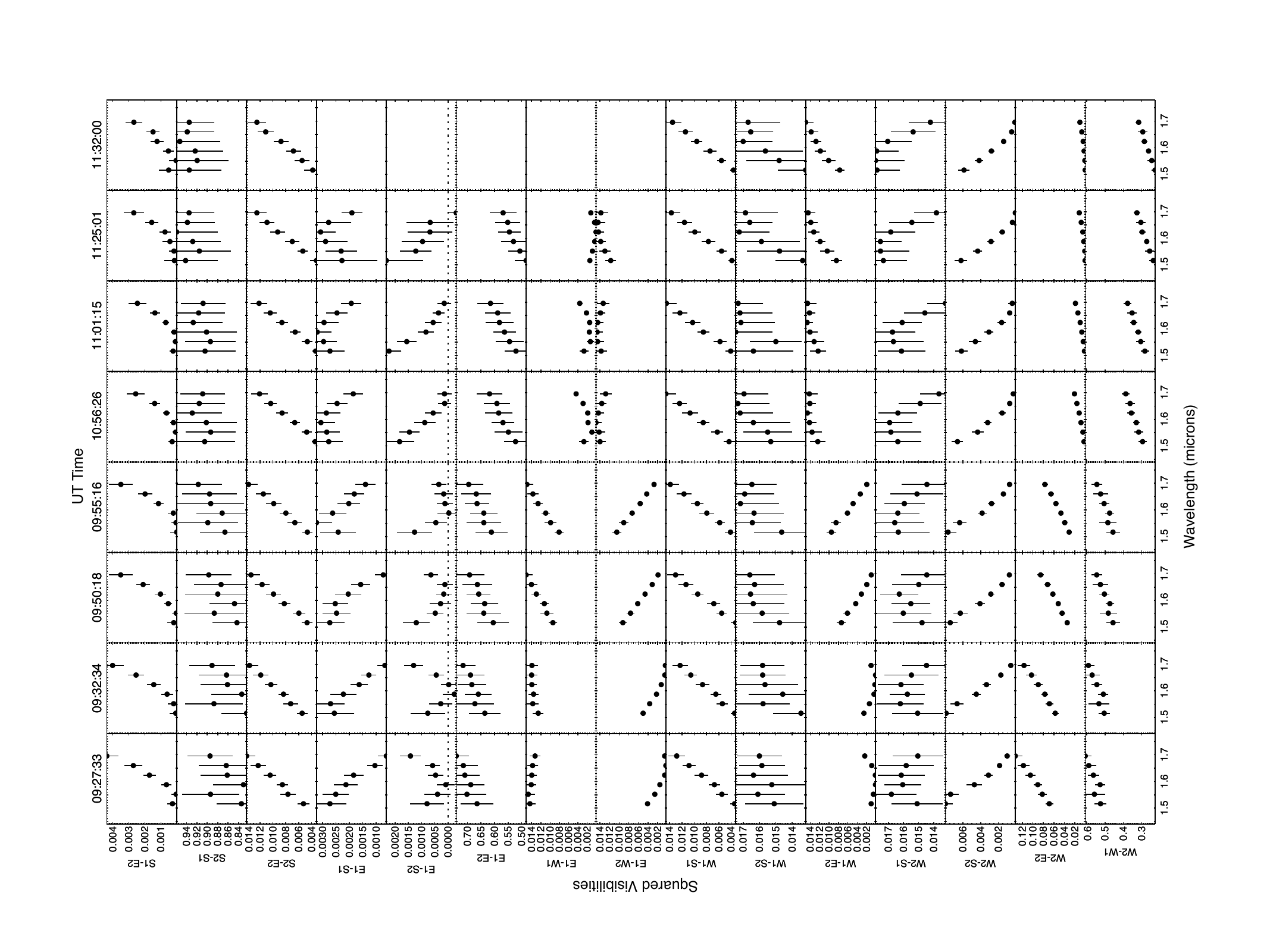}
    \caption{Squared visibilities of UT 2019 July 17 $\zeta$~And CHARA/MIRC-X observations with $1\sigma$ error bars.  Each block represents a pair of telescopes in a temporal block of observations.   
    } 
    \label{fig:v2block}
\end{center}
\end{figure*}

\begin{figure*}
\begin{center}
    \includegraphics[angle=270,trim={1.5cm 1.5cm 1.5cm 1.5cm},clip,width=0.9\textwidth]{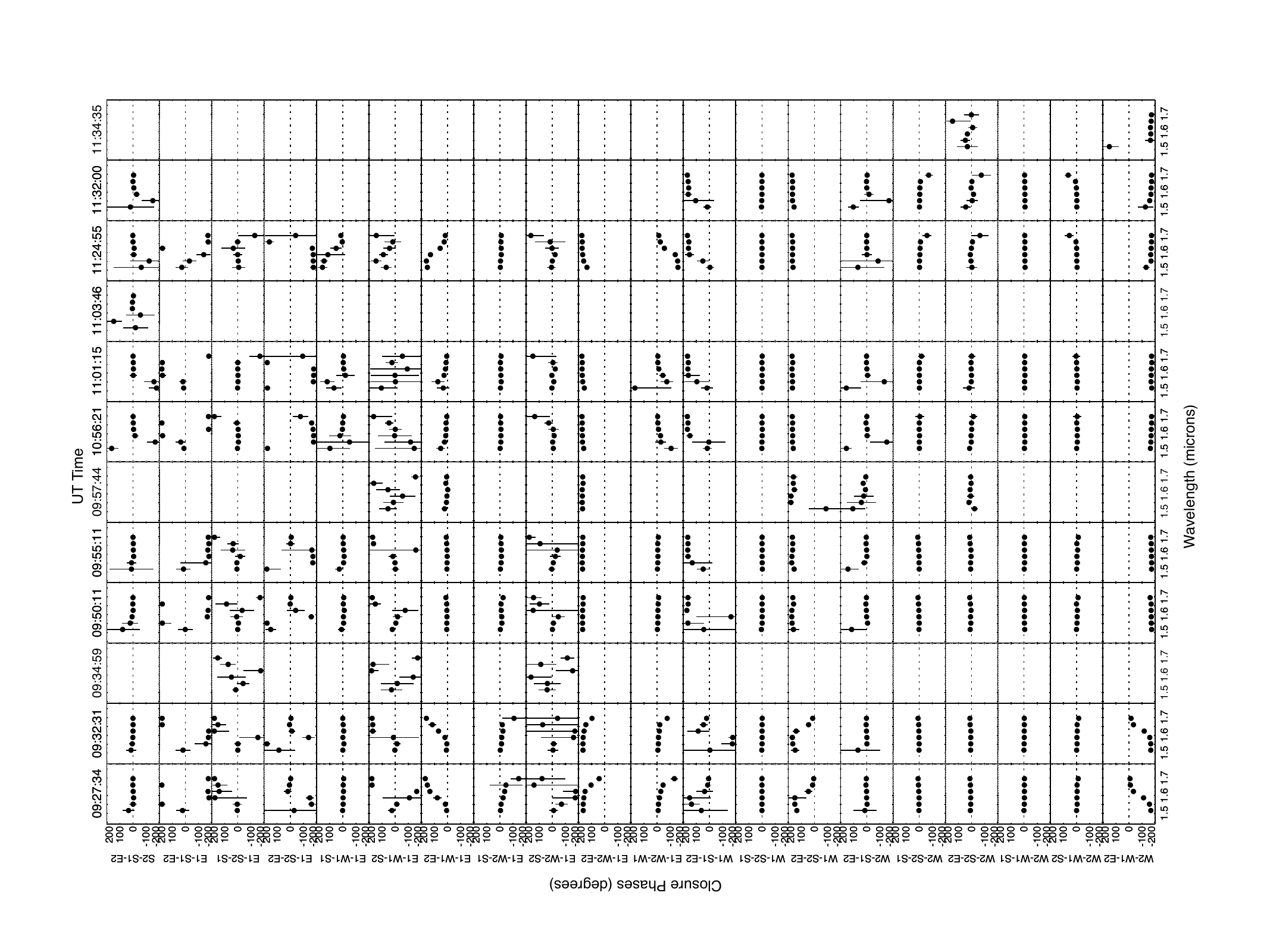}
    \caption{Closure phases of UT 2019 July 17 $\zeta$~And CHARA/MIRC-X observations with $1\sigma$ error bars.  Each block represents a set of three telescopes in a temporal block of observations.   
    } 
    \label{fig:cpblock}
\end{center}
\end{figure*}

\begin{figure*}
\begin{center}
    \includegraphics[angle=270,trim={1.5cm 1.5cm 1.5cm 1.5cm},clip,width=0.9\textwidth]{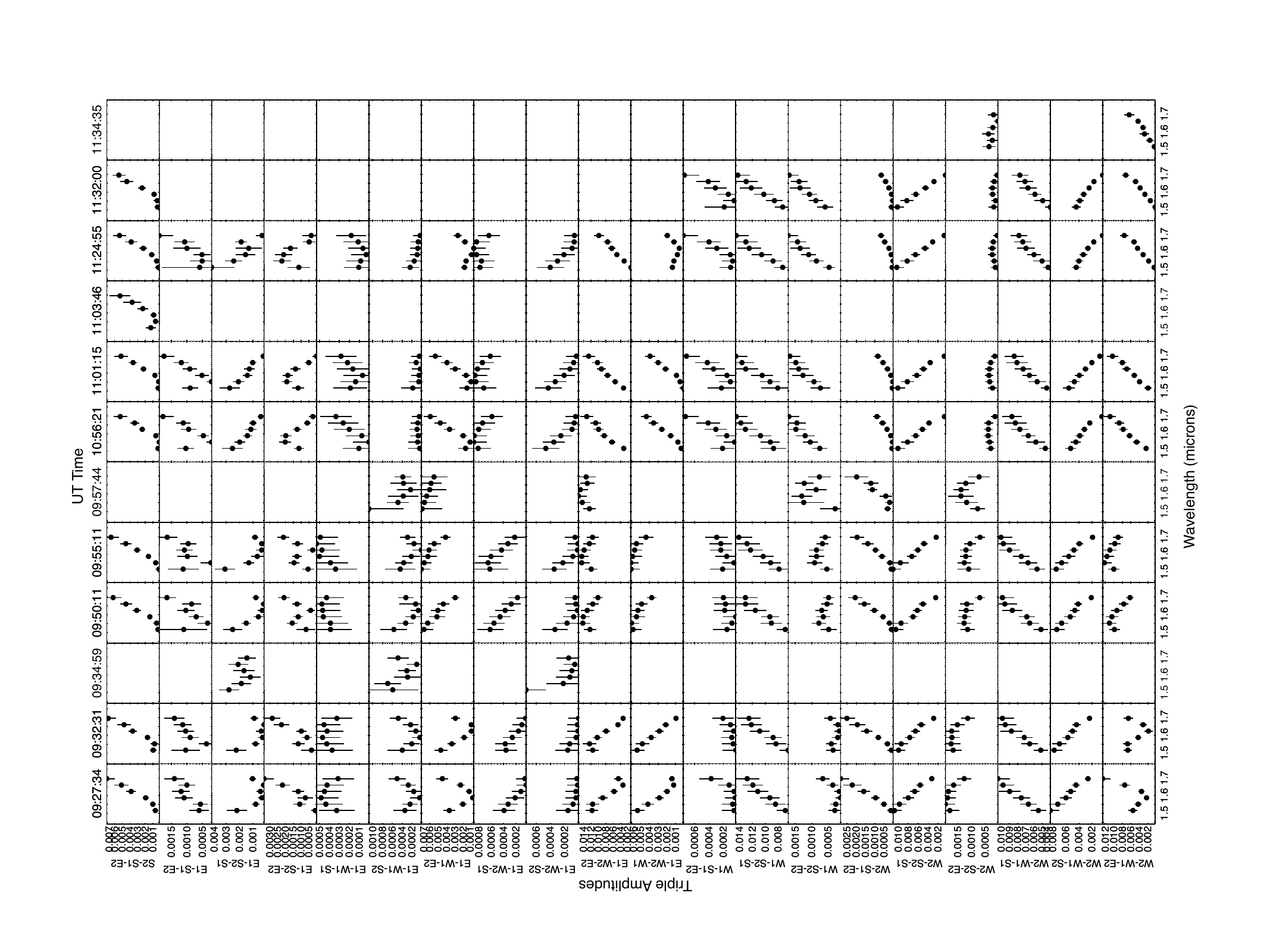}
    \caption{Triple amplitudes of UT 2019 July 17 $\zeta$~And CHARA/MIRC-X observations with $1\sigma$ error bars.  Each block represents a set of three telescopes in a temporal block of observations.   
    } 
    \label{fig:T3ampblock}
\end{center}
\end{figure*}

The data were reduced with the MIRC-X pipeline (version 1.3.3, jdm-develop branch) and were calibrated using software originally developed for the MIRC beam combiner data \citep{monnier12}, but modified for MIRC-X data.  The data were median filtered over wavelength.  Five stars were used for calibration, and they are listed in Table \ref{Table:MIRCXobs} for each night they were used.  Table \ref{Table:Cals} contains information on the calibration stars. 
 Example observations are shown in Figures \ref{fig:viscurve}--\ref{fig:T3ampblock}.

\begin{figure}
    \includegraphics[trim={2.5cm 2.5cm 2.5cm 12cm},clip,width=0.475\textwidth]{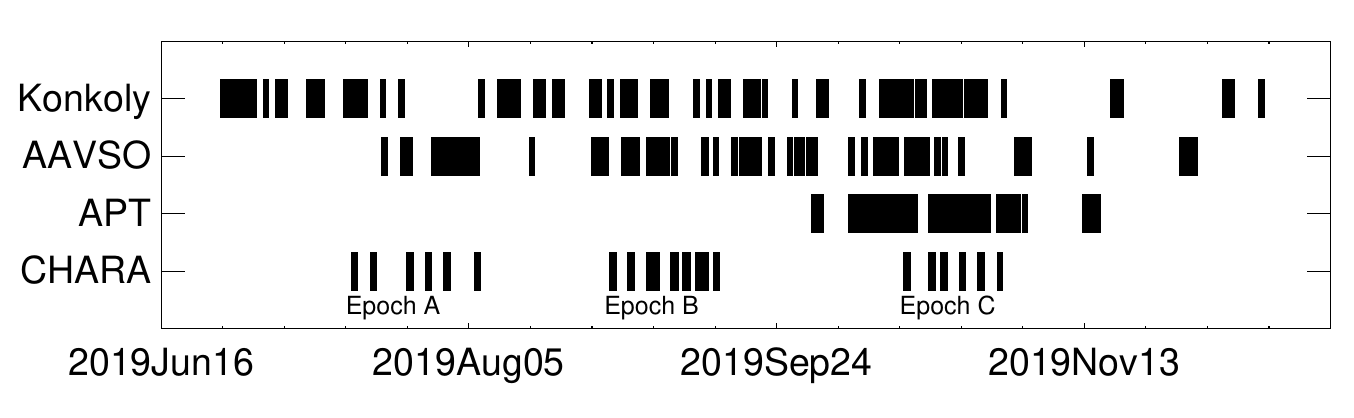}
    \caption{Timeline representation of the interferometric and photometric observations.  The CHARA Array observations are described in Section \ref{section:mircx}.  The APT photometry are described in Section \ref{section:apt}, the AAVSO photometry in Section \ref{section:aavso}, and the Konkoly Observatory photometry in Section \ref{section:konkoly}.}
    \label{fig:timeline}
\end{figure}

Data were obtained in three epochs spanning 105 nights:  2019 July 17 - August 6 (Epoch A), 2019 August 28 - September 14 (Epoch B), and 2019 October 15 - 30 (Epoch C).  Epoch A spans 20.00 days (1.13 rotations) and includes data from six nights of observation.  Epoch B spans 16.99 days (0.96 rotations) and includes data from nine nights of observation.  Epoch C spans 15.17 days (0.85 rotations) and includes data from six nights of observation.  Between Epochs A and B are 21.87 days (1.23 rotations).  Between Epochs B and C are 30.87 days (1.74 rotations).  The specific dates and information on the telescopes used and the number of observations are included in Table \ref{Table:MIRCXobs} with a timeline in Figure \ref{fig:timeline}.
 
\begin{deluxetable*}{l c c c c}
\tabletypesize{\scriptsize}
\tablecaption{MIRC-X Observations}
\tablewidth{0pt}
\tablehead{
\colhead{UT Date} &  \colhead{Telescopes Used} & \colhead{Number of $V^2$}  & \colhead{Number of Closure Phases} &
\colhead{Calibration Stars Used}
}
\startdata
 \multicolumn{5}{c}{ \textbf{Epoch A}}  \\
  \hline
2019 July 17 & E1-W2-W1-S2-S1-E2 & 690 & 972 & HD 14055, HD 219080 \\
2019 July 20 & E1-W2-W1-S2-S1-E2 & 546 & 738 & HD 886,  HD 14055, HD 219080\\ 
2019 July 26 & E1-W2-W1-S2-S1-E2 & 852 & 116 & HD 886,  HD 14055, HD 219080 \\
2019 July 29 & E1-W2-W1-S2-S1-E2 & 816 & 1164 & HD 886,  HD 14055, HD 219080\\
2019 August 1 & E1-W2-W1-S2-S1-E2 & 558  & 858 & HD 886,  HD 14055, HD 219080 \\
2019 August 6 & E1-W2-W1-S2-S1-E2 &  264 & 354 &  HD 886\\
  \hline
 \multicolumn{5}{c}{ \textbf{Epoch B}}  \\
 \hline
 2019 August 28 & E1-W2-W1-S2-S1-E2 & 960 & 1308 & HD 886,  HD 14055, HD 219080, HD 220657 \\
 2019 August 31 & E1-W2-W1-S2-S1-E2 & 1254 & 1728 & HD 886, HD 6903,  HD 14055, HD 219080\\
 2019 September 3 & E1-W2-W1-S2-S1-E2 & 882 & 1062 & HD 886,  HD 14055, HD 220657\\
 2019 September 4 & W2-W1-S2-S1-E2 & 444 & 486 & HD 886, HD 14055, HD 220657 \\
 2019 September 7 & E1-W2-W1-S2-S1-E2 & 1332 & 1848 & HD 886, HD 6903,  HD 14055, HD 219080\\
 2019 September 9 & E1-W2-W1-S2-S1-E2 & 1022 & 1057 & HD 886, HD 6903,  HD 14055, HD 219080\\
 2019 September 11 & E1-W2-W1-S2-S1-E2 & 120 & 168 & HD 219080\\
 2019 September 12 & E1-W2-W1-S2-S1-E2 & 1284 & 1788 & HD 886,  HD 14055, HD 219080, HD 220657\\
 2019 September 14 & E1-W1-S2-S1-E2 & 126 & 144 & HD 219080 \\ 
  \hline
  \multicolumn{5}{c}{ \textbf{Epoch C}}  \\
 \hline
 2019 October 15 & E1-W2-W1-S2-S1-E2 & 726 & 1044 & HD 886, HD 6903,  HD 14055, HD 219080\\ 
 2019 October 19 & E1-W2-W1-S2-S1-E2 & 756 & 1134 & HD 886, HD 6903,  HD 14055, HD 219080\\ 
 2019 October 21 & E1-W2-W1-S2-S1-E2 & 918 & 1326 & HD 886, HD 6903,  HD 14055, HD 219080\\ 
 2019 October 24 & E1-W2-W1-S2-S1-E2 & 906 & 1326 & HD 886, HD 6903,  HD 14055, HD 219080\\ 
 2019 October 27 & E1-W2-W1-S2-S1-E2 & 828 & 1194 & HD 886, HD 6903,  HD 14055, HD 219080\\ 
 2019 October 30 & E1-W2-W1-S2-S1-E2 & 354 & 528 & HD 6903 \\ 
\enddata
  \tablecomments{See Table \ref{Table:Cals} for details on calibration stars. }
\label{Table:MIRCXobs}
\end{deluxetable*}

\begin{deluxetable}{l c c c}
\tabletypesize{\scriptsize}
\tablecaption{MIRC-X Calibration Stars}
\tablewidth{0pt}
\tablehead{
\colhead{HD Number} & \colhead{Common Name} & \colhead{$H$} & \colhead{Uniform Disk } \\
\colhead{} &  \colhead{} & \colhead{Magnitude} & \colhead{ Angular Diameter}  \\
\colhead{} &  \colhead{} & \colhead{} &  \colhead{(mas)}  
}
\startdata
 HD 886 & $\gamma$ Peg & 3.638 & $0.372 \pm 0.038$ \\
 HD 6903 & $\psi^{3}$ Psc & 4.189 & $0.606 \pm 0.071$ \\ 
 HD 14055 & $\gamma$ Tri &  3.862 & $0.549 \pm 0.051$ \\
 HD 219080 & 7 And &  3.760 & $0.693 \pm 0.068$ \\ 
 HD 220657 & $\upsilon$ Peg & 3.230 & $0.973 \pm  0.088$ \\
\enddata
  \tablecomments{Angular diameters and errors are from \citet{chelli16}.  $H$~magnitudes are from \citet{cutri03}. }
\label{Table:Cals}
\end{deluxetable}

\subsection{Automatic Photoelectric Telescope Photometry}
\label{section:apt}

$\zeta$~And was observed with the Tennessee State University T3 0.4m Automatic Photoelectric Telescope (APT) located at Fairborn Observatory in the Patagonia Mountains of southern Arizona.  Johnson $B$ and $V$ data were collected from 2019 September 29 through November 14, overlapping with Epoch C.  Monsoon season in Arizona prevented contemporaneous observations with all of the interferometric data.  

The T3 APT uses a single-channel photometer equipped with a temperature-stabilized EMI 9924B photomultiplier tube that measures count rates through the $B$ and $V$ filters.  Each differential photometric observation of $\zeta$~And consists of $B$ and $V$ measurements in the following variable star  group sequence:  K, S, C, V, C, V, C, V, C, S, K, where K is the check star  (HD 6009), C is the comparison star (HD 5286), V is $\zeta$~And, and S is a  sky reading.  From each observational sequence, three V-C and two K-C differential magnitudes are created in each passband and averaged together to create $B$ and $V$ group means.  Group mean differential magnitudes with internal standard deviations greater than 0.01 mag are discarded to eliminate observations taken under nonphotometric conditions.  The surviving group means were corrected for differential extinction, transformed to the standard Johnson photometric system, and treated as single observations thereafter.  This group sequence was repeated up to six times each clear night at intervals of 2-3 hr.  The reduced differential magnitudes are found in Table~\ref{Table:APT}.  For further information on the operation of the T3 APT and reduction of the data, see \citet{henry1999} and \citet{fekel2005}.  

\begin{figure}
    \includegraphics[trim={2.5cm 5 2.5 5},clip,width=0.5\textwidth]{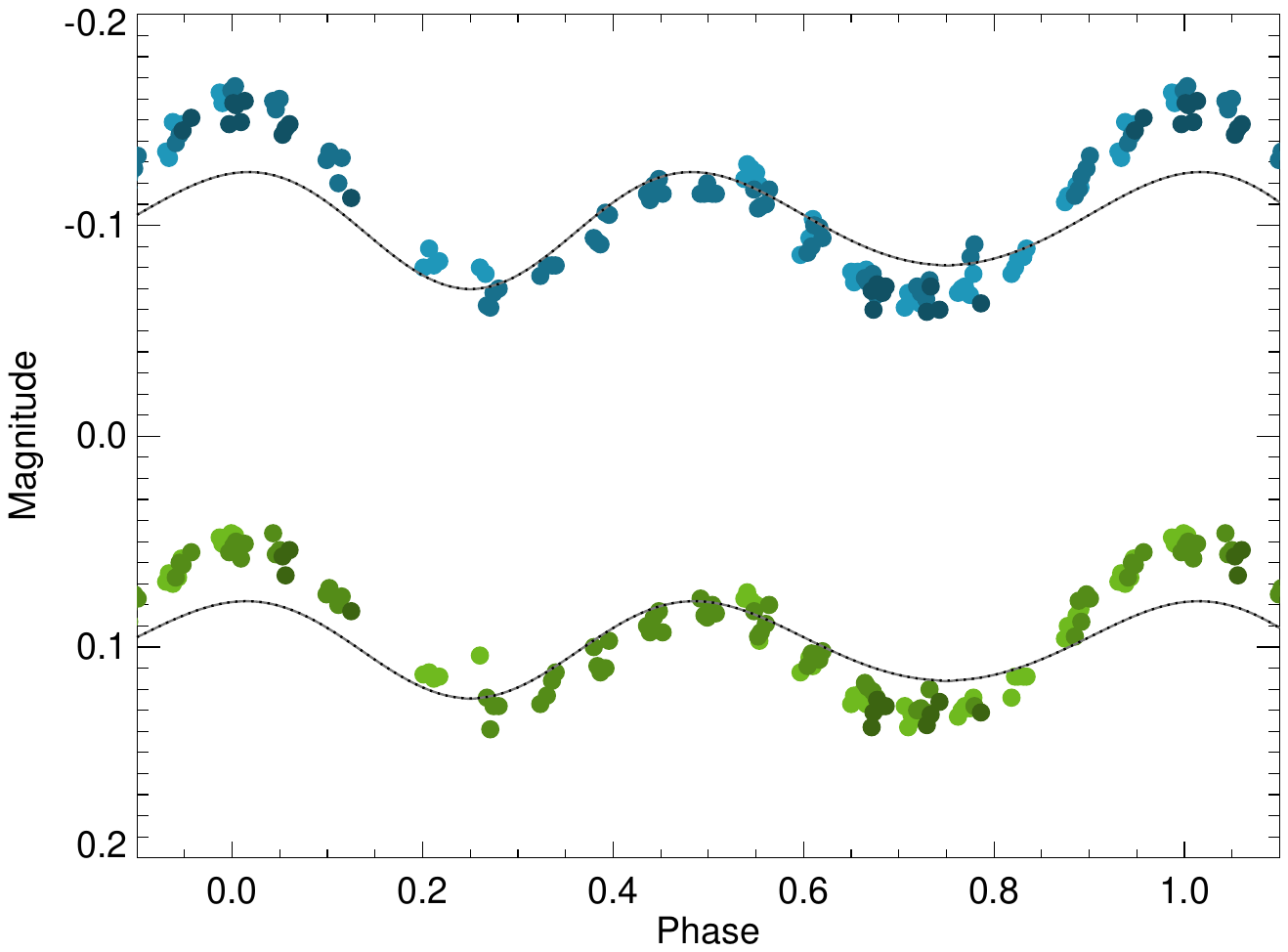}
    \caption{$B$- and $V$-band APT light curves from 2019 September 29 - November 14.  The shade of blue ($B$-band) and green ($V$-band) depends upon the rotation of the star with the shade darkening as time progresses, consistent with the shading used in Figures \ref{fig:AAVSO} and \ref{fig:Konkoly}.  The gray line for each bandpass represents the appropriate ELC-modeled light curve for $\zeta$~And with a main-sequence companion ($0.75 \ M_\odot$, $0.80 \ R_\odot$, and $4800$~K; ELC is discussed in Section \ref{section:surfacefeatures}).  The black dashed line superimposed on the gray line represents the ELC model with a white dwarf companion  ($0.75 \ M_\odot$, $0.01 \ R_\odot$, and $25000$~K).  
    The lines are nearly identical, except for slight deviations at phases 0.25 and 0.75 where the main-sequence model has very shallow eclipses that would not be detectable with our photometry.   The light curves are arbitrarily shifted in magnitude to be plotted together. The mismatch between the ELC models and the observed light curve is attributed to the presence of starspots.
    } 
    \label{fig:APTELC}
\end{figure}

Adopting the $P_\mathrm{rot} = P_\mathrm{orb} = 17.769426 \pm 0.000040$ days and the time of nodal passage $T_0 = 49992.281 \pm 0.017$ (MJD)\footnote{The time of nodal passage is defined as the point in the orbit when the primary is at its maximum positive velocity.} from \citep{fekel1999}, the phase-folded $\zeta$~And APT light curves are shown in Figure \ref{fig:APTELC}.

\begin{deluxetable*}{l c c c c}
\tabletypesize{\scriptsize}
\tablecaption{Johnson $B$ and $V$ APT Photometry}
\tablewidth{0pt}
\tablehead{
\colhead{Reduced Julian date } &  \colhead{$\Delta B$ Magnitude}  & \colhead{$\Delta V$ Magnitude } &  \colhead{$\Delta B$ Magnitude } & \colhead{$\Delta V$ Magnitude}\\
\colhead{} & \colhead{($\zeta$~And - Comp)} & \colhead{($\zeta$~And - Comp)} & \colhead{(Check - Comp)} & \colhead{(Check - Comp)}
}
\startdata
58756.6785	&	$-1.244$	&	$-1.351$	&	1.018	&	1.244	\\
58756.7823	&	$-1.253$	&	$-1.352$	&	1.018	&	1.247	\\
58756.8699	&	$-1.245$	&	$-1.349$	&	1.023	&	1.244	\\
58756.9720	&	$-1.247$	&	$-1.350$	&	1.015	&	1.247	\\
58757.7307	&	$-1.244$	&	$-1.360$	&	--	&	--	\\
... & ... & ... & ... & ... \\
\enddata
\tablecomments{This table includes a sample of the data.  The full data set is available as a machine-readable table.  }
\label{Table:APT}
\end{deluxetable*}

\subsection{AAVSO Photometry}
\label{section:aavso}

$\zeta$~And was observed through the American Association of Variable Star Observers \citep[AAVSO;][]{kloppenborg25}.  Bright Star Monitor telescopes New Mexico (65-mm aperture) and New Hampshire II (60-mm aperture) collected Johnson $B$, $V$, and $I$ data from 2019 July 30 through November 30.  These observations overlapped partially with Epoch A of the interferometric data and completely with Epochs B and C.  

\begin{figure}
    \includegraphics[trim={2.5cm 5 2.5 5},clip,width=0.5\textwidth]{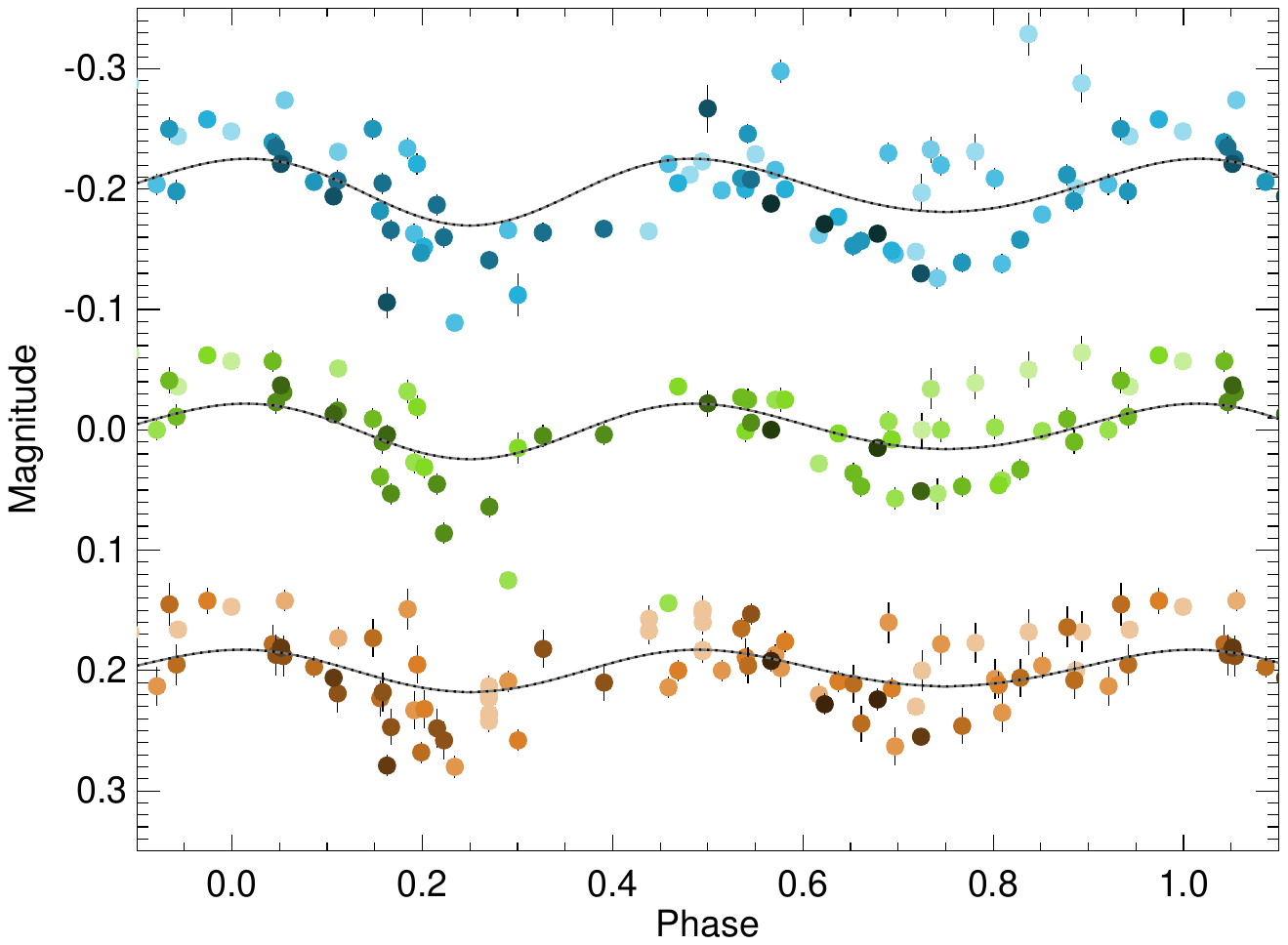}
    \caption{$B$-, $V$-, and $I$-band light curves from AAVSO spanning 2019 July 30 - November 30.  The shade of blue ($B$-band), green ($V$-band), and orange ($I$-band) depends upon the rotation of the star with the shade darkening as time progresses, as in Figure \ref{fig:APTELC}.  The ELC-modeled light curves follow the same convention as in Figure \ref{fig:APTELC} with the gray line representing a main-sequence companion and the black dashed line represents a white dwarf companion. The light curves are arbitrarily shifted in magnitude to be plotted together. 
    } 
    \label{fig:AAVSO}
\end{figure}

Our observations are differential measurements with comparison stars  HD 4832, HD 4269, and HD 4105.  The data are included in Table \ref{Table:AAVSO}, and the phase-folded AAVSO light curves are shown in Figure \ref{fig:AAVSO}.

\begin{deluxetable*}{l c c c c}
\tabletypesize{\scriptsize}
\tablecaption{Johnson $BVI$ AAVSO Photometry}
\tablewidth{0pt}
\tablehead{
\colhead{Heliocentric Julian Date } &  \colhead{Filter} & \colhead{Magnitude}  & \colhead{ Magnitude Error} & \colhead{Telescope}
\\
}
\startdata
2458689.811623	& $B$	&5.158	&0.004	&BSM\textunderscore NH2\\
2458690.579767	& $B$	&5.111	&0.004	&BSM\textunderscore NH2\\
2458690.810150	& $B$	&5.100	&0.004	&BSM\textunderscore NH2\\
2458691.806167	& $B$ &5.094	&0.006	&BSM\textunderscore NH2\\
2458694.797819	& $B$	&5.175	&0.005	&BSM\textunderscore NH2\\
... & ... & ... & ... & ... \\
\hline
2458694.911505 & $V$ & 4.092 & 0.014 & BSM\textunderscore NM \\
2458695.911796 & $V$ & 4.053 & 0.014 & BSM\textunderscore NM \\
2458696.905098 & $V$ & 4.042 & 0.015 & BSM\textunderscore NM \\
2458697.898730 & $V$ & 4.028 & 0.014 & BSM\textunderscore NM \\
2458698.790369& $V$ & 4.003 &	0.005 &	BSM\textunderscore NH2 \\
... & ... & ... & ... & ... \\
\hline
2458686.826077	& $I$ &	2.893	 & 0.009	 & BSM\textunderscore NH2 \\
2458686.827710	& $I$ &	2.896	 & 0.009	  & BSM\textunderscore NH2\\ 
2458686.829345	& $I$ &	2.915	 &	0.009 & 	BSM\textunderscore NH2 \\
2458686.830970	& $I$ &	2.886 & 	0.009 & 	BSM\textunderscore NH2 \\
2458686.832568	& $I$ &	2.909 & 	0.009	 & BSM\textunderscore NH2  \\
... & ... & ... & ... & ... \\
\enddata
\tablecomments{BSM\textunderscore NM is the Bright Star Monitor telescope in New Mexico.  BSM\textunderscore NH2 is the Bright Star Monitor telescope New Hampshire II.  This table includes a sample of the data.  The full data set is available as a machine-readable table.  }
\label{Table:AAVSO}
\end{deluxetable*}

\subsection{Konkoly Observatory Photometry}
\label{section:konkoly}

$\zeta$~And was observed with the 0.6\,m Schmidt telescope at the Piszk\'estet\H o Mountain Station of Konkoly Observatory in Hungary.  Johnson $B$, $V$, $R$, and $I$ data were collected from 2019 June 26 through December 11, overlapping all epochs of our interferometric data. The following objects were used as comparison stars: TYC~1739-712-1 (HD 4373), TYC~1739-649-1, TYC~1739-548-1, and TYC~1739-1090-1. Aperture photometry was performed with the \texttt{fitsh} package \citep{2012MNRAS.421.1825P}, following bias subtraction and flat-field correction. At each night at least 3 $BVRI$ sequences were made, and the final light curve was created by taking daily medians.

\begin{figure}
    \includegraphics[trim={2.5cm 6 2.5 6},clip,width=0.5\textwidth]{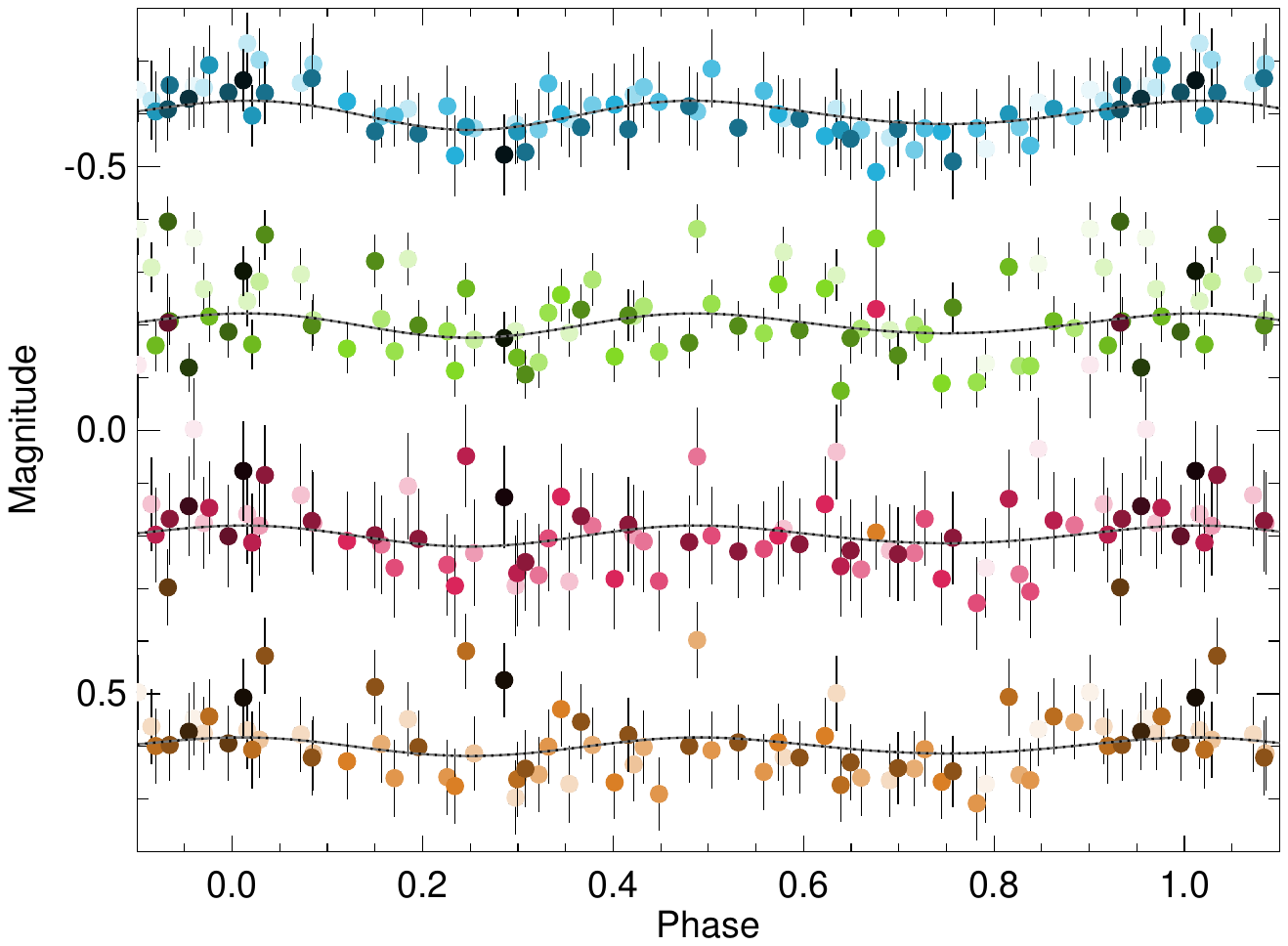}
    \caption{$B$-, $V$-, $R$-, and $I$-band light curves from Konkoly Observatory spanning 2019 June 26 - December 11.  The shade of blue ($B$-band), green ($V$-band), red ($R$-band), and orange ($I$-band) depends upon the rotation of the star with the shade darkening as time progresses, as in Figure \ref{fig:APTELC}.  The ELC-modeled light curves follow the same convention as in Figure \ref{fig:APTELC} with the gray line representing a main-sequence companion and the black dashed line represents a white dwarf companion.   The light curves are arbitrarily shifted in magnitude to be plotted together. 
    } 
    \label{fig:Konkoly}
\end{figure}

The data are included in Table \ref{Table:konkoly} with the phase-folded Konkoly Observatory light curves are shown in Figure \ref{fig:Konkoly}.

\begin{deluxetable*}{l c c c c c c c c}
\tabletypesize{\scriptsize}
\tablecaption{Johnson $BVRI$ Konkoly Observatory Photometry}
\tablewidth{0pt}
\tablehead{
\colhead{Julian Date } &  \colhead{$B$ Magnitude}  & \colhead{$B$ Magnitude} &  \colhead{$V$ Magnitude}  & \colhead{$V$ Magnitude} & \colhead{$R$ Magnitude}  & \colhead{$R$ Magnitude} & \colhead{$I$ Magnitude}  & \colhead{$I$ Magnitude } \\
\colhead{} & \colhead{} & \colhead{Error} & \colhead{} & \colhead{Error} & \colhead{} & \colhead{Error} & \colhead{} & \colhead{Error}
}
\startdata
2458660.55549 & 5.289 & 0.059 & 4.366 & 0.047 & 3.904 & 0.094 & 3.316 & 0.073 \\
2458661.53379 & 5.201 & 0.077 & 4.177 & 0.049 & 3.678 & 0.097 & 3.212 & 0.071 \\
2458662.49857 & 5.177 & 0.061 & 4.111 & 0.050 & 3.519 & 0.101 & 3.142 & 0.071 \\
2458663.54370 & 5.170 & 0.077 & 4.128 & 0.049 & 3.641 & 0.096 & 3.192 & 0.070 \\
2458664.54120 & 5.089 & 0.074 & 4.248 & 0.046 & 3.802 & 0.093 & 3.214 & 0.073 \\
... & ... & ... & ... & ... & ... & ... & ... & ... \\
\enddata
\tablecomments{This table includes a sample of the data.  The full data set is available as a machine-readable table.  }
\label{Table:konkoly}
\end{deluxetable*}

\subsection{TESS Photometry}

\begin{figure}
    \includegraphics[trim={2.5cm 5 2.5 5},clip,width=0.5\textwidth]{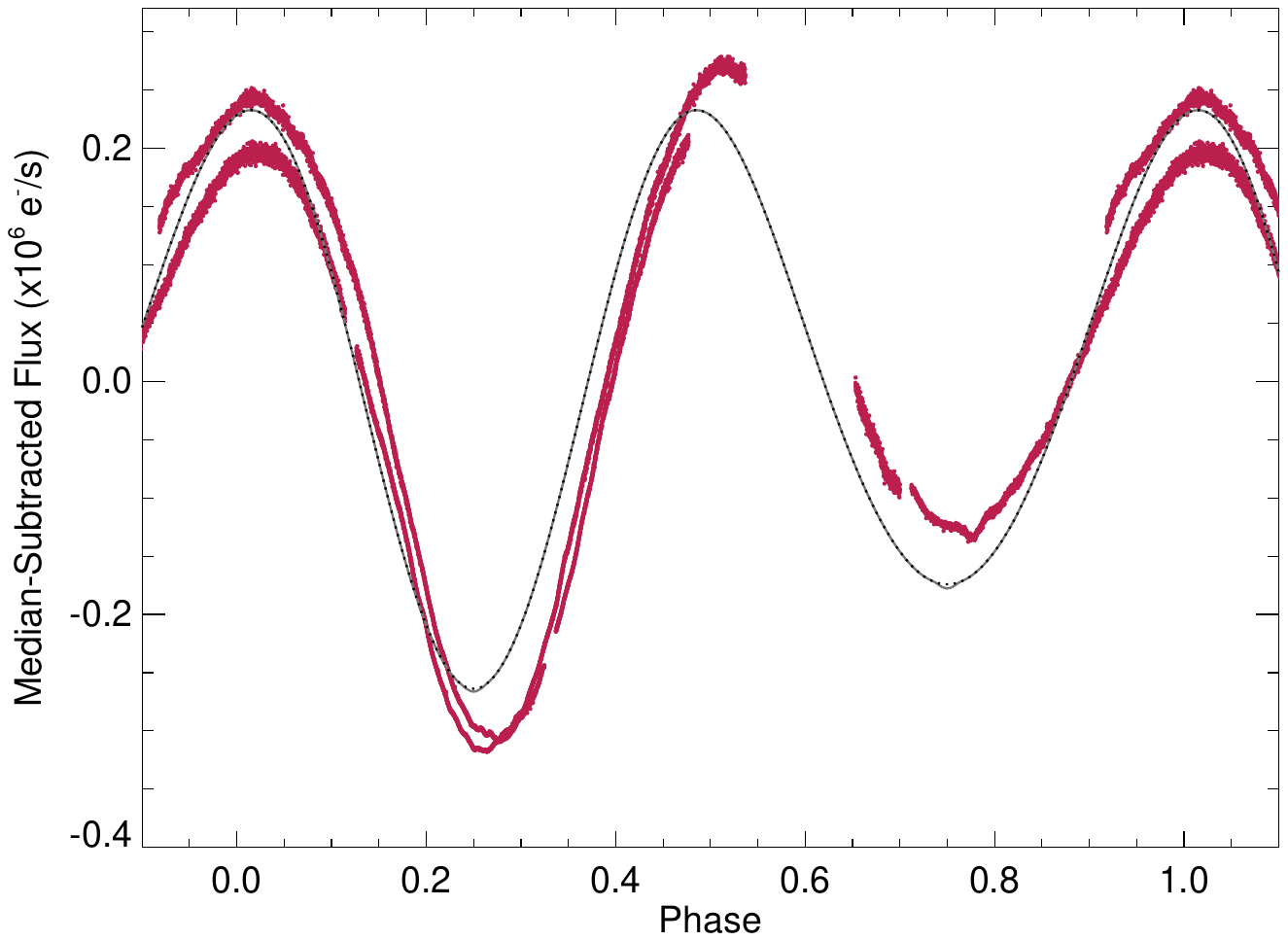}
    \caption{Non-contemporaneous TESS light curve from Sector 57 (2022 July 9--August 5; median-subtracted and in red) compared to modeled light curves from ELC.  The ELC-modeled light curves follow the same convention as in Figure \ref{fig:APTELC} with the gray line representing a main-sequence companion and the black dashed line represents a white dwarf companion. The ELC models are arbitrarily scaled to roughly match the magnitude of the TESS data for comparison.  The TESS light curve shows no evidence of an eclipse, as seen in the ELC model with the main-sequence companion.  
    } 
    \label{fig:TESS}
\end{figure}

$\zeta$~And was observed by the Transiting Exoplanet Survey Satellite \citep[TESS;][]{ricker14} in Sectors 17 and 57.  Sector 17 spanned 2019 October 7 through November 2.  While these data were contemporaneous with our interferometric observations, $\zeta$~And fell on the edge of the detector and the motion of the satellite made the Sector 17 light curve unusable.  Sector 57 occurred from 2022 September 30 to October 29.  We retrieved the data using standard TESS pipeline from the Barbara A. Mikulski Archive for Space Telescopes (MAST).  We minimally processed the simple aperture photometry (SAP) light curve only by removing points flagged as bad (nonzero quality flag).  The phase-folded Sector 57 data are shown in Figure \ref{fig:TESS}.

\

\section{Interferometric Imaging}

Aperture synthesis imaging was performed on our interferometric data sets to reveal the structures of the stellar surface.  We use the SURFace imagING \citep[SURFING;][]{roettenbacher16b} algorithm to combine the data from an individual stellar rotation and image the star as a whole.  In \citet{martinez21}, a data set of the spotted star $\lambda$~And was imaged with both SURFING and  another imaging algorithm, ROTational Image Reconstruction (ROTIR).  This test of the algorithms showed good agreement in the results, with only minor differences in the starspot locations and contrast.

The SURFING algorithm assumes that the surface does not change significantly during each epoch, i.e., there is neither differential rotation nor significant starspot evolution over the course of our observed epochs.  The image from SURFING will be an average of any overlapping data that show evolution between the nights of observation.

Following \citet{roettenbacher16b},  we fix three  parameters in SURFING.  The rotation period is set to the orbital period determined from radial velocity studies \citep[$P_\mathrm{rot} = P_\mathrm{orb} = 17.769426\pm0.000040$~days;][]{fekel1999}.   The time of nodal passage was also taken from \citet{fekel1999}, $T_0 = 49992.281 \pm 0.017$ (MJD).  Because of the difficulty in measuring the limb darkening when the surface is very spotted due to confusion between limb darkening and spots on the limb, we defined our power-law limb-darkening coefficient to be 0.269, after \citet{roettenbacher16b}. 

To determine the remaining stellar parameters of $\zeta$~And needed for SURFING---limb-darkened stellar angular diameter, $\theta_\mathrm{LD}$; visibility at the origin where the spatial frequency is 0.0), $V_0$; oblateness of the star; angle of inclination, $i$; and position angle of the rotational pole (East of North)---we combine all observations from the three epochs.  We first allowed a global fit (the free parameters were allowed to take on any value) with SURFING on a low-resolution stellar surface (126 resolution elements) to restrict these parameters (800 walkers with 1000 iterations).  With the results of the global fit as priors, we performed 50 instances (each running 24 walkers over 500 iterations) of SURFING on all 2019 data together on a surface with 768 elements, matching the CHARA Array's spatial resolution.  A surface element covers an area of $0.025$~mas$^2$.  To achieve a $\chi^2 \sim 1.0$, we multiplied the error bars on the interferometric data by 1.65.  We report the mean and $1\sigma$ errors of each parameter in Table \ref{Table:SURFINGparams}.  In Table \ref{Table:SURFINGparams}, the limb-darkened angular diameter resulting from SURFING was divided by a wavelength correction factor.  We applied a  correction  of $1.0054 \pm 0.0006$ to obtain an angular diameter of $\theta_\mathrm{LD}=2.495 \pm 0.014$ mas.  The wavelength correction factor  follows from a required scaling that was found by \citet{gardner22} when combining data from MIRC-X with data from the GRAVITY instrument at the Very Large Telescope Interferometer.  The MIRC-X data used an internally consistent wavelength calibration, but the GRAVITY data use absolute calibration.  The scaling value was updated by J.\ D.\ Monnier (private communication).

\begin{figure*}
\begin{center}
    \includegraphics[scale = 0.3]{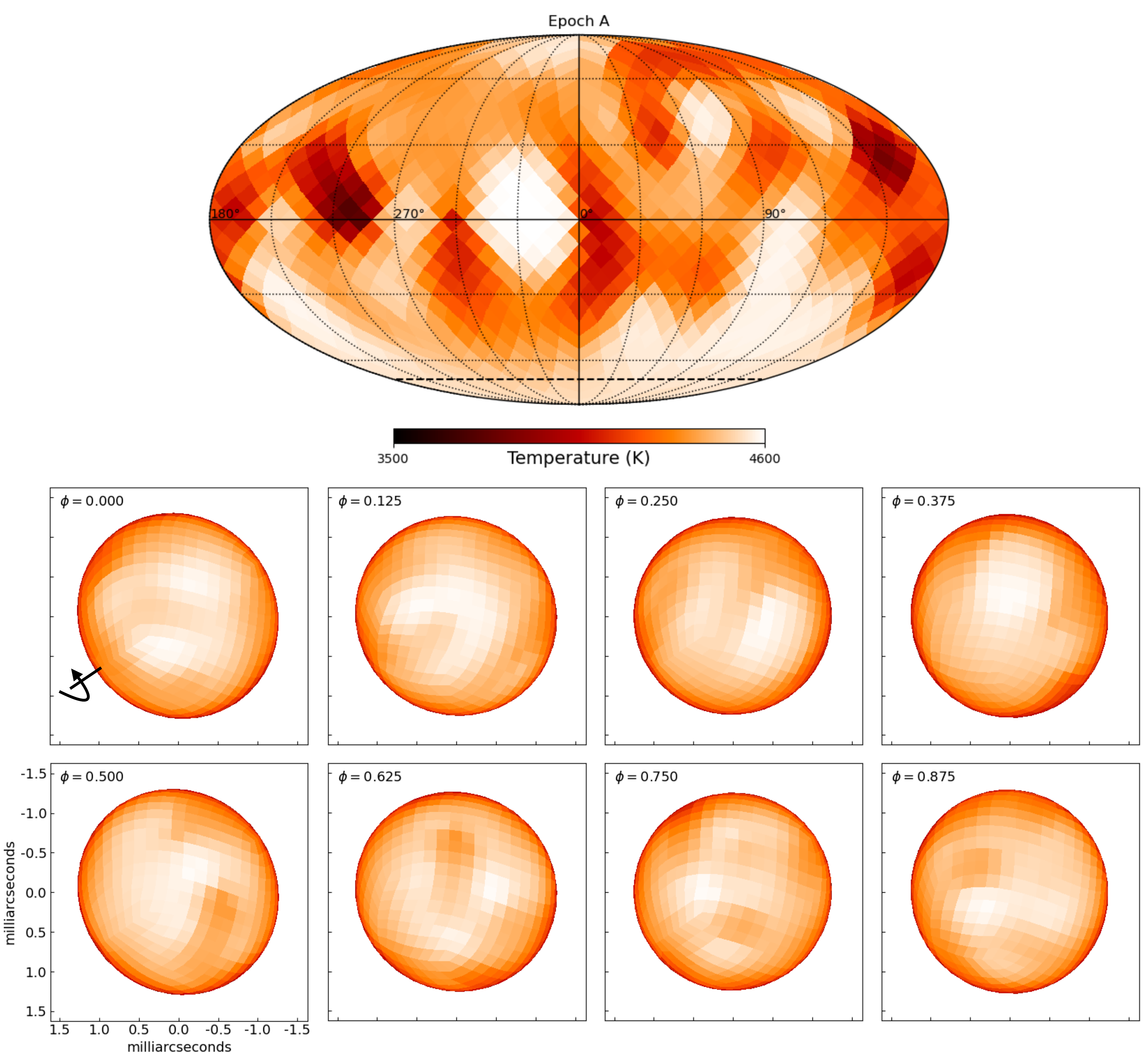}
    \caption{Top: Mollweide projections of the temperature of $\zeta$~And using data from Epoch A.  The dashed line at bottom marks the latitude below which the stellar surface is not observable due to the stellar inclination.  Bottom:  Orthographic projections of $\zeta$~And as it appeared on the sky during Epoch A in $H$-band.
    } 
    \label{fig:zetAndA}
\end{center}
\end{figure*}

\begin{figure*}
\begin{center}
    \includegraphics[scale = 0.3]{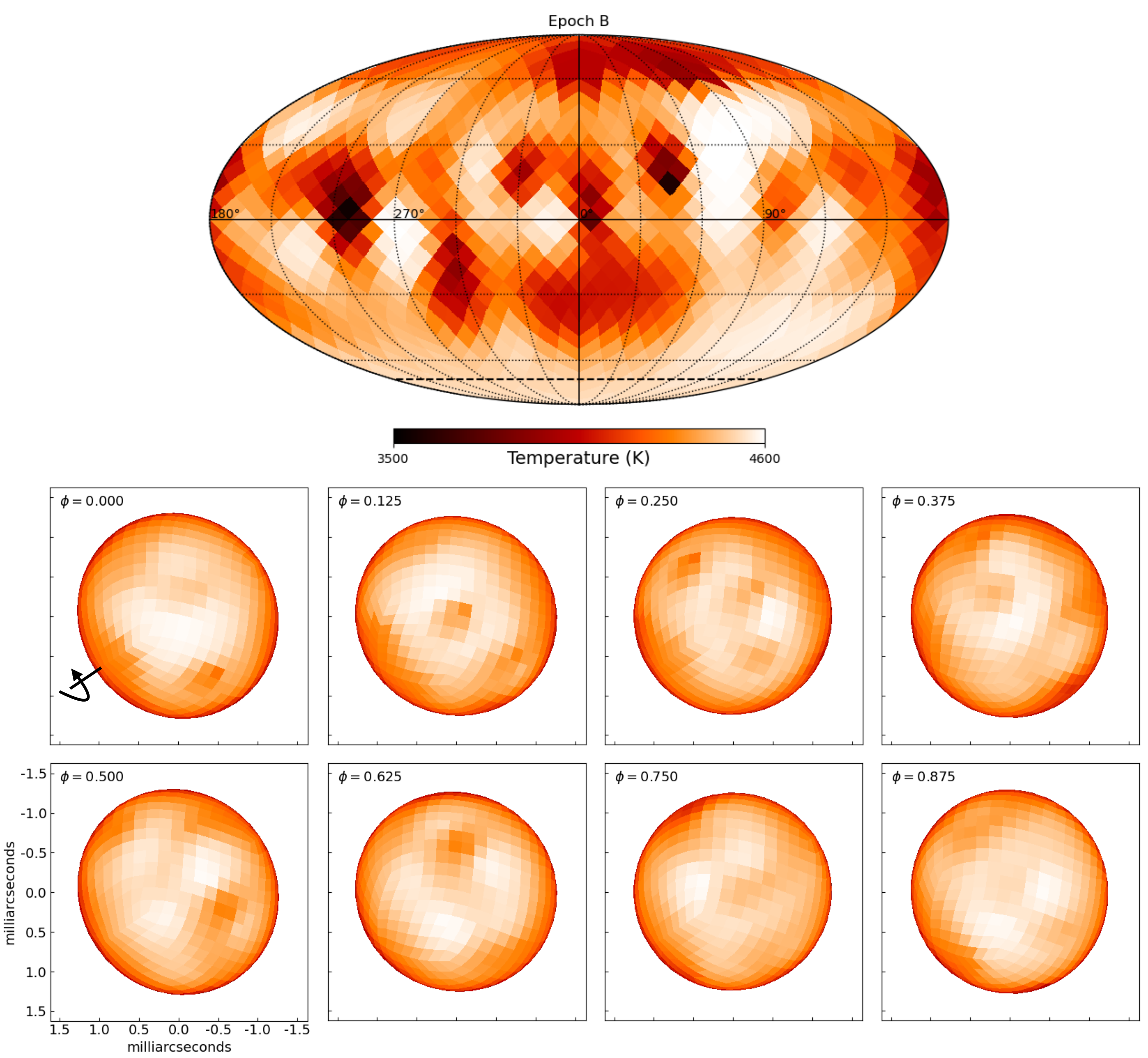}
    \caption{Top: Mollweide projections of the temperature of $\zeta$~And using data from Epoch B.  The dashed line at bottom  marks the latitude below which the stellar surface is not observable due to the stellar inclination.  Bottom:  Orthographic projections of $\zeta$~And as it appeared on the sky during Epoch B in $H$-band.
    } 
    \label{fig:zetAndB}
\end{center}
\end{figure*}

\begin{figure*}
\begin{center}
    \includegraphics[scale = 0.3]{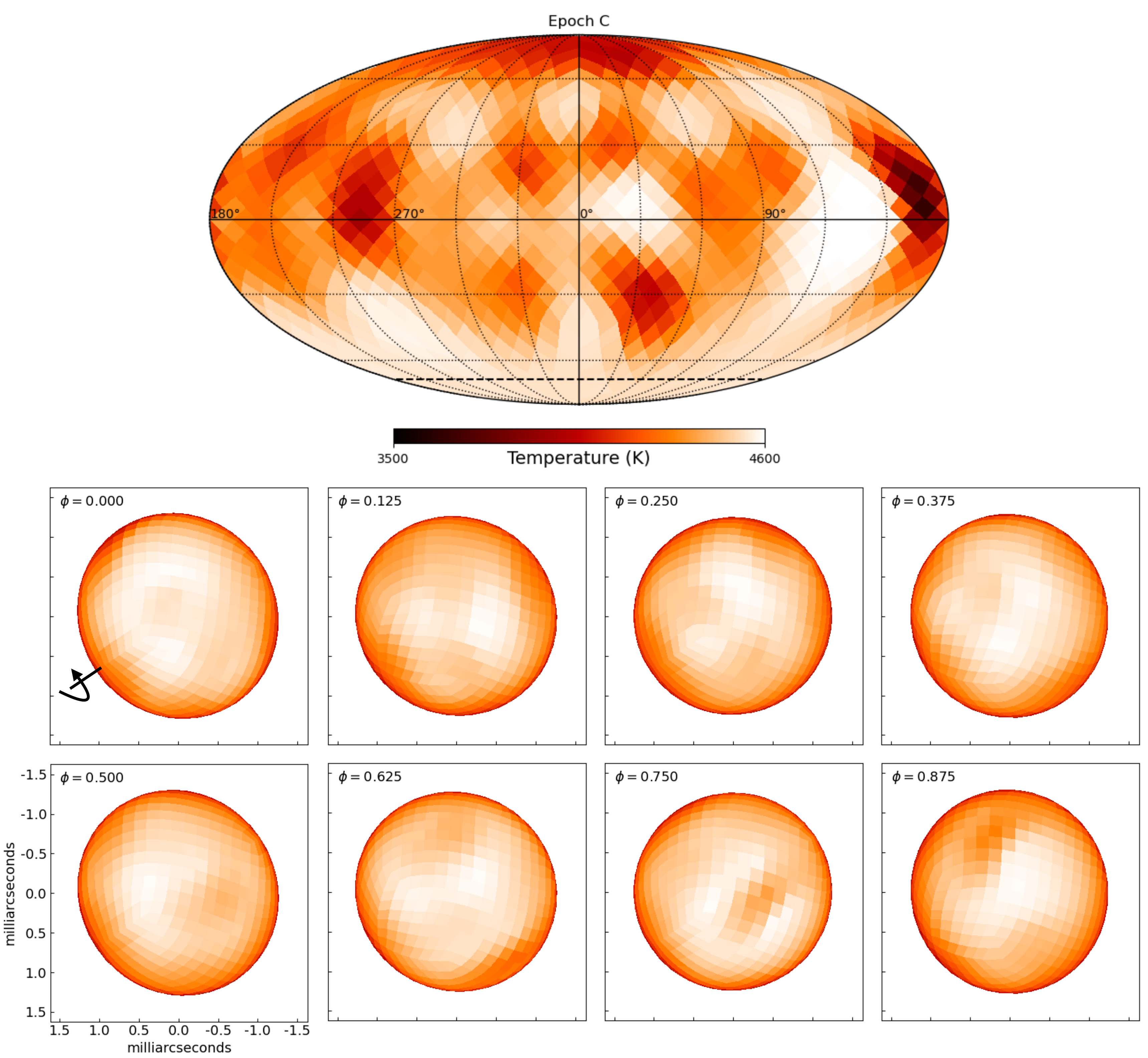}
    \caption{Top: Mollweide projections of the temperature of $\zeta$~And using data from Epoch C.  The dashed line at bottom marks the latitude below which the stellar surface is not observable due to the stellar inclination.  Bottom:  Orthographic projections of $\zeta$~And as it appeared on the sky during Epoch Cin $H$-band.
    } 
    \label{fig:zetAndC}
\end{center}
\end{figure*}

These values were then used as the input parameters for SURFING for the individual epochs.  The resultant images are shown Figures \ref{fig:zetAndA}--\ref{fig:zetAndC}. To achieve $\chi^2 \approx 1.0$, error bars on the interferometric data were multiplied by 1.60 for Epoch A, 1.40 for Epoch B, and 1.25 for Epoch C.  Without these multiplicative factors, the $\chi^2$ values are all above 1.  These factors aim to ensure that the data are not being under- or over-fit and do not impact the nature of the spotted surface in the resultant images.  Each resultant image is an average of 50 instances (6 walkers over 100 iterations) fixing the stellar parameters and allowing only the surface brightness  of each patch to vary.

\begin{deluxetable*}{l c c}
\tabletypesize{\scriptsize}
\tablecaption{SURFING Stellar Parameters}
\tablewidth{0pt}
\tablehead{
\colhead{Parameter } &  \colhead{Value  } & \colhead{Value } \\
\colhead{} & \colhead{\citep{roettenbacher16b}} & \colhead{(this work)}
}
\startdata
Limb-darkened disk diameter, $\theta_\mathrm{LD}$ (mas) & $2.502 \pm 0.008$ & $2.495 \pm 0.014$ \\
Limb-darkening coefficient (power law), $\alpha$ & 0.269 & 0.269 \\
Visibility at the origin, $V_0$ & $1.00 \pm 0.02$ & $1.02 \pm 0.02$ \\
Major-to-minor axis ratio & $1.060 \pm 0.011$ & $1.063 \pm 0.013$\\
Inclination, $i$ ($^\circ)$ & $70.0 \pm 2.8$ &  $69.3 \pm 6.9$ \\
Rotational pole position angle, E of N ($^\circ$) & $126.0 \pm 1.9$ & $124.0 \pm 2.9$\\
Rotation period, $P_\mathrm{rot}$ (days) & 17.769426 & 17.769426 \\
Time of nodal passage, $T_0$ (MJD) & 49992.281 & 49992.281 \\
\enddata
\tablecomments{Those values not shown with error bars are fixed input parameters.}
\label{Table:SURFINGparams}
\end{deluxetable*}

\subsection{Stellar Surface Features}
\label{section:surfacefeatures}

Figures \ref{fig:zetAndA}--\ref{fig:zetAndC} show the stellar surface of $\zeta$~And at each observed epoch. The top of each figure shows a Mollweide projection of the the star's temperature map, while the bottom shows the intensity of the star in $H$-band as it appeared on the sky. During our observations, the star is covered in large starspots, including a polar spot.  While some spots may be traced from one epoch to the next, the surface evolution between the epochs is substantial, which makes definitively following the spots between the epochs difficult due to the gaps in observation.  This is in contrast to archival long-term photometry \citep{strassmeier89,kaye95,strassmeier1999c,zhang2000,kovari07,korhonen10}, which suggests that the surface of $\zeta$~And evolves slowly, retaining the same shape (dominated by the star's ellipsoidal variations) over the course of months or years.  The photometry that we include here from the APT, AAVSO, and Konkoly Observatory were all obtained contemporaneously to our interferometric observations (as demonstrated in Figure \ref{fig:timeline}).  Like the archival data, these observations show the light curve is dominated by the star's ellipsoidal shape with small deviations from that due to starspots.  During the photometric observations, the shape of the light curve does not change dramatically, and much of the variation seen is due to scatter in the data. 

\begin{figure}
\begin{center}
    \includegraphics[trim={2.5cm 5 2.5 5},clip,width=0.5\textwidth]{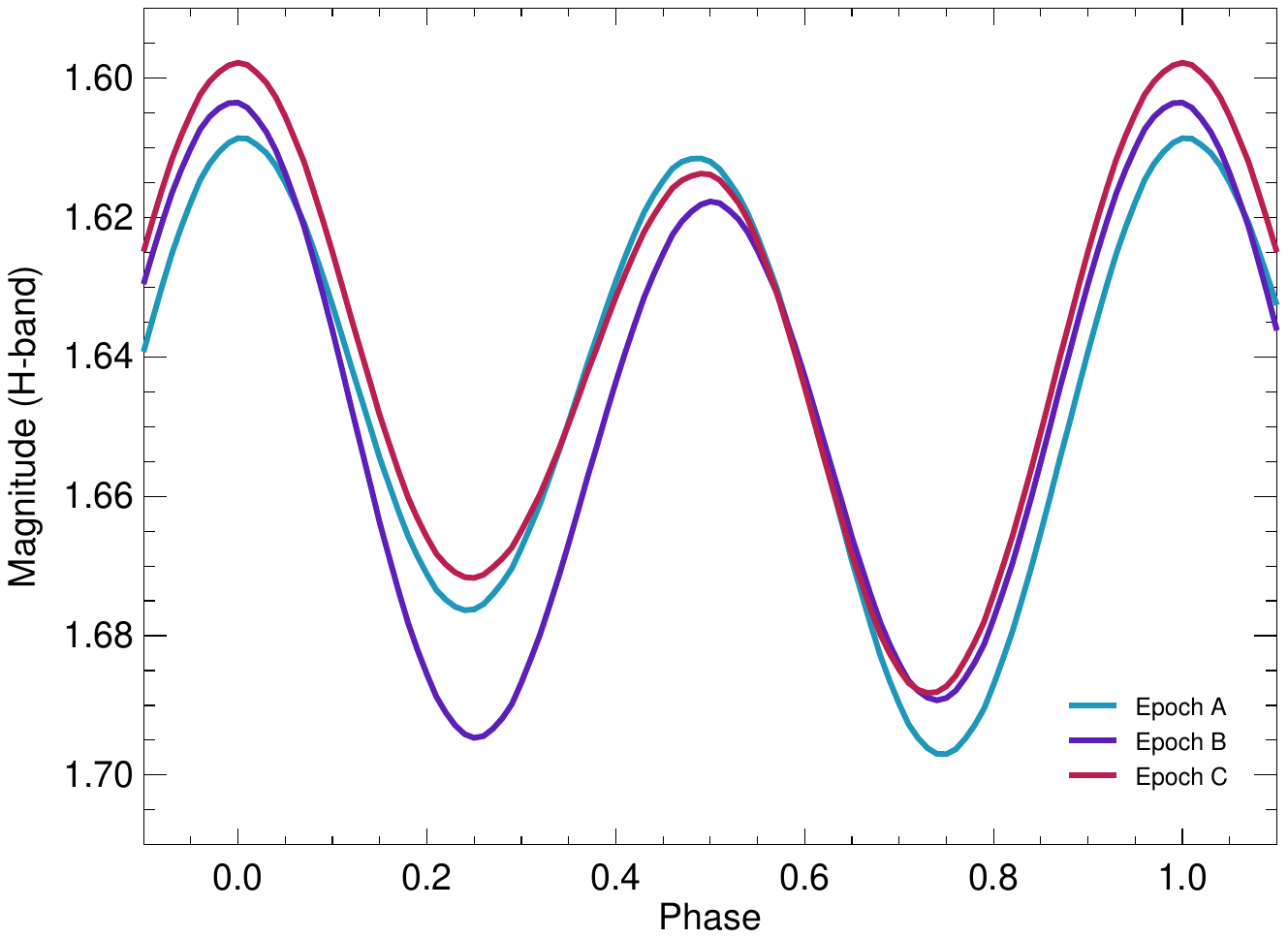}
    \caption{$H$-band light curves calculated from the SURFING images.  The overall shape of each light curve is very similar, but there are small changes in the amplitude of variation from the evolving starspots.  
    } 
    \label{fig:SURFINGLCs}
\end{center}
\end{figure}

To demonstrate that the stellar surfaces we are presenting from the interferometric data are consistent with photometry, we modeled the resultant light curves from each image.  These light curves are shown in Figure \ref{fig:SURFINGLCs}.  As with the observed photometry, the overall shape of the light curve is dominated by the ellipsoidal variation signature.  The changing surface features do impact the light curves, but at a scale that is comparable to the scatter of the observations.  Additionally, we modeled the ellipsoidal signature of the system with the fitting and modeling algorithm Eclipsing Light Curve \citep[ELC;][]{orosz2000}.  Based upon the known stellar parameters of $\zeta$~And \citep[from][]{kovari07,roettenbacher16b} and the parameters determined here, ELC produced a model light curve for the geometry and limb darkening of $\zeta$ And without starspots.  We include model light curves on Figures \ref{fig:APTELC} and \ref{fig:AAVSO}.  The light curves are largely characterized by the ELC model.  The ELC model does not perfectly match the observations because of the starspots present on the actual surface of $\zeta$ And that are not included in the ELC model.

\begin{figure}
\begin{center}
    \includegraphics[trim={2.5cm 5 2.5 5},clip,width=0.5\textwidth]{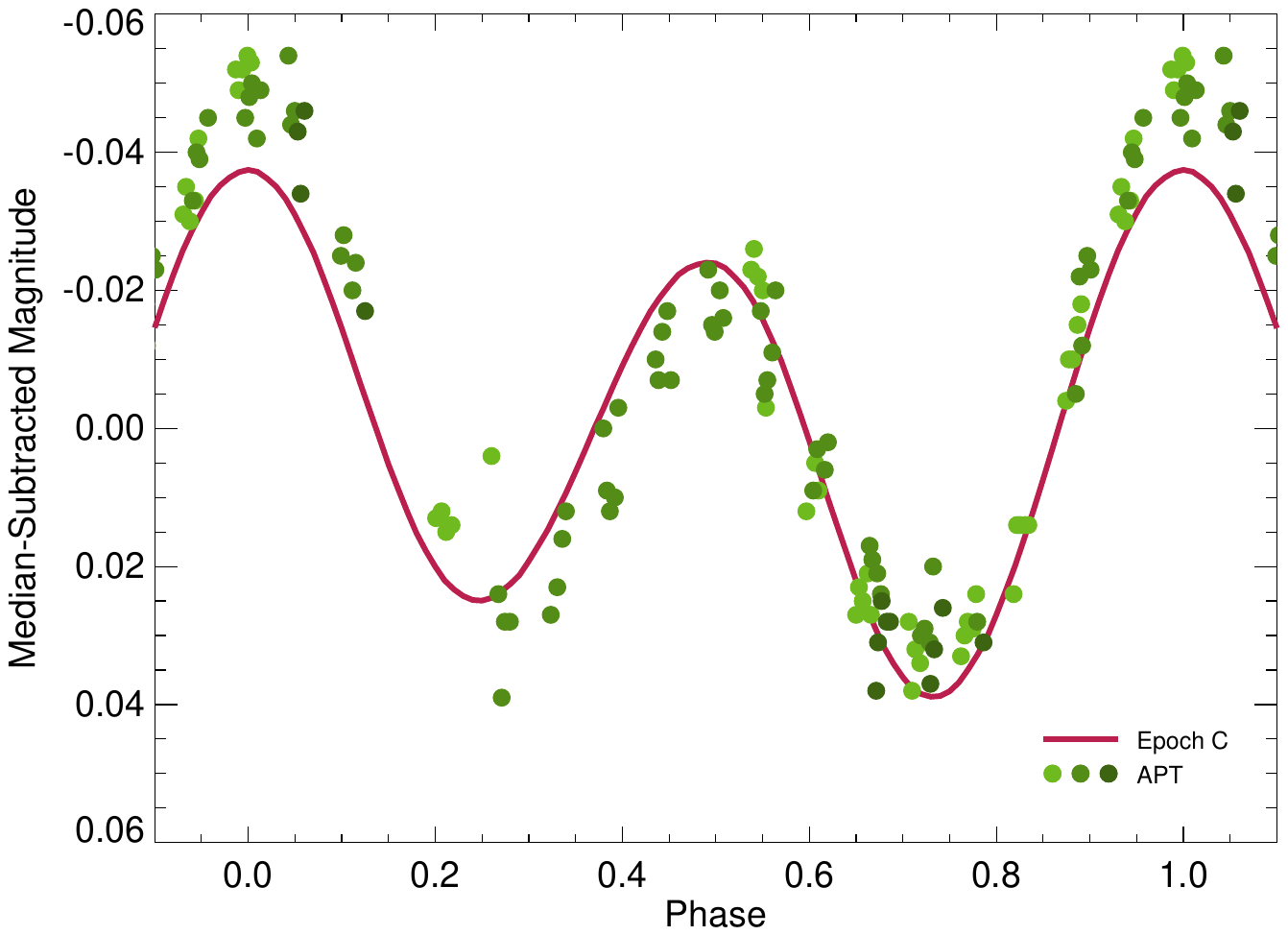}
    \caption{Comparison of the contemporaneous median-subtracted APT $V$-band light curve (green dots) and the SURFING light curve from Epoch C (red line) transformed to $V$-band.  The shade of the green for the APT data depends upon the rotation of the star.  The shade darkens as time progresses.
    } 
    \label{fig:APTSURFING}
\end{center}
\end{figure}

We directly compare the APT data to the modeled light curve from SURFING for Epoch C in Figure \ref{fig:APTSURFING}, as the observations occurred at the same time.  We interpolated between the $V$- and $H$-band based on the ELC models. The shape of the observed light curve is well-matched by the light curve generated by the SURFING image.  

The interferometric images of $\zeta$~And from 2011 and 2013 included in \citet{roettenbacher16b} similarly show large spots distributed across the entire stellar surface.  Both of those earlier epochs show a persistent polar spot, which is an indication that the magnetic dynamo generating the starspots of $\zeta$~And differs from that of the Sun.  In \citet{roettenbacher16b}, the polar spot was among the coolest regions on the stellar surface.  In the images here, the polar spot is not among the coolest features on the stellar surface.  In fact, the polar spot appears to grow in size across our new images, suggesting the feature may be transient, like the spots at lower latitudes.  As seen in \citet{roettenbacher16b}, the starspots appear at all latitudes.  There is no evidence of particularly active latitudes, which are seen on the Sun \citep[e.g.,][]{hathaway2015}.
In the 2013 observations, a spot is present on the equator at $180^\circ$ longitude.  In the 2019 images here, a spot is present near this longitude on the equator in each epoch.  When this longitude is facing Earth, the orbit is at phase = 0.75; at this phase, the primary star is at the point in its orbit where it is closest to Earth and the secondary is furthest.  Our sample of images is not large enough to conclude if the point on the star  opposite of the companion is an active region or if the spot locations are coincidental.

\subsection{Differential Rotation}

Three epochs of observations allow for an investigation into the differential rotation of $\zeta$~And.   Efforts to track stellar spots to measure differential rotation have previously made use of Doppler imaging \citep{kovari07,kovari2012}.  In \citet{kovari2012}, sequential Doppler images were cross-correlated to measure the extent of differential rotation.  They found that the star had solar-like differential rotation (equatorial regions rotating more rapidly than polar regions). 
We cannot apply the same methods here, as the starspots are not moving uniformly from one image to the next.  In fact, the starspots are not drifting with respect to one another as expected if due to differential rotation.    

\citet{kovari2012} gave a solar-type differential rotation law of 
\[ \Omega (\beta) = 20.689 (1 - 0.055 \sin^2 \beta ), \]
where $\Omega$ is the rate of rotation (given as $^\circ$/day) at the specific latitude $\beta$ (in $^\circ$).  
This differential rotation law is based on observations spanning about 49 days (our observations span approximately 105 days).  To illustrate the magnitude of the \citet{kovari2012} differential rotation law, consider the dark spot on the equator in all three of our epochs seen at approximately a longitude of $240^\circ$ and the dark spot near $-40^\circ$ latitude and $300^\circ$ longitude in Figure \ref{fig:zetAndA}.  Because the equator is assumed to be rotating at the same rate that the companion is orbiting due to tidal locking, the rotation period assigned to $\zeta$~And is the rotation period of its equator.  We expect the equatorial spot to remain at about the same longitude from rotation to rotation.  With respect to that spot, the spot at $-40^\circ$ latitude is expected to advance $0.47^\circ$ in longitude fewer per day, based on the \citet{kovari2012} differential rotation law.  For one rotation period, this amounts to a longitudinal shift of about $8.4^\circ$.  From Epoch A to Epoch C, this spot should shift $41^\circ$ (nearly 4 pixels) in longitude further away from the equatorial spot.  
The lower-latitude spot appears to have drifted about the same amount of longitude as the equatorial spot.  The spot's evolution is likely, as the spot's temperature shown in Figures \ref{fig:zetAndA}--\ref{fig:zetAndC} changes and may be interacting with other nearby spots as time progresses.  As a result of the evolution and the gap in time between the epochs, we are unable to definitively trace the starspots over time.    

Giant stars are expected to have weaker differential rotation than that of the Sun \citep{henry1995}.  While the \citet{kovari2012}  differential rotation law is weaker than what is seen on the Sun, our images suggest that if $\zeta$~And is differentially rotating, its law must actually be even weaker.  
Based on our images, the timescale of spot evolution appears to occur faster than spot motion due to the star's differential rotation.   The evolution of the starspots could also be what was detected by \citet{kovari2012} and misinterpreted as differential rotation.

\section{The Nature of the Companion}
\label{section:companion}

\citet{kovari07} reported a mass of $M_1 = 2.6 \pm 0.4 M_\odot$ for the primary star based upon stellar evolution models, which  is consistent with past studies \citep[e.g.,][]{gratton1950,stawikowski1994,fekel1999}. Using the mass ratio $q = M_2/M_1 = 0.29$, as reported by \citet{gratton1950} and \citet{stawikowski1994}, \citet{kovari07}, gives $M_2 \approx 0.75 \ M_\odot$.  All past spectroscopic observations identified $\zeta$~And as a single-lined spectroscopic binary, with the companion undetected, likely due to the brightness of the primary star overwhelming the light of the companion \citep{campbell1911,cannon1915,SpencerJones1928,gratton1950,hendry1980,beavers1986,fekel1999,massarotti2008}.  The companion has been assumed to be a main sequence star, as is most  common with other RS CVn systems.  

With our interferometric data, we performed a search for a binary companion. 
Should the companion of $\zeta$~And have been a main-sequence star, we would expect a flux ratio of the primary star to the secondary star in $H$-band to be $\approx 300:1$ \citep[secondary mass of $0.75 \ M_\odot$ from][and estimated $R = 0.8 \ R_\odot$ and $T_\mathrm{eff} = 4800$~K]{kovari07}.  This is a brightness ratio that would be easily detected by the combination of MIRC-X and the CHARA Array, which has detected flux ratios up to nearly 1300:1 \citep{evans2024}.  For similar RS CVn systems, flux ratios between the primary and the companion of 370:1 and 270:1 have been detected \citep{roettenbacher15a,roettenbacher15b} with the predecessor beam combiner, MIRC \citep{monnier04}.  These past successes suggest that a contrast ratio of 300:1 would not go undetected in our large dataset.  If it is the case that the companion does have a mass of $\approx 0.75 \ M_\odot$, and its contrast ratio is not within the detection limits of MIRC-X and the CHARA Array, the companion could potentially be a white dwarf.

To further investigate the nature of the companion, we estimated the secondary's temperature and radius based on whether the $\approx0.75 \ M_\odot$ star is a single star on the main sequence ($R = 0.8 \ R_\odot$, $T_\mathrm{eff} = 4800 \ K$) or a white dwarf ($R = 0.01 \ R_\odot$, $T_\mathrm{eff} = 25000 \ K$).  We modeled the light curve for the system with ELC using the known orbital and stellar parameters of the primary star of $\zeta$~And.  As expected by ELC, we used bolometric and wavelength-specific limb-darkening coefficients from \citet{vanhamme1993}.  With the inclination of $69.3^\circ$, a single main-sequence companion would eclipse the primary star, but the amplitude of the eclipse is under $\Delta V =0.001$ magnitudes, which is well below the errors of our photometry and unlikely to be detected.  If the actual of angle of inclination were $76.2^\circ$, $1\sigma$  away from the best-fit inclination of $69.3^\circ$, the primary eclipse would have a magnitude of $\Delta V =0.005$ magnitudes.  This size eclipse begins to be detectable with some ground-based resources, and would be readily detected in space-based photometric data.  The white dwarf companion does not produce a detectable eclipse, regardless of inclination.   
Light curves of the binary system with both a main-sequence star and a white dwarf as the companion are shown in the phase-folded  light curves in Figures \ref{fig:APTELC} and \ref{fig:AAVSO} (the best-fit inclination is assumed for the plotted models).  

We include a non-contemporaneous TESS light curve in Figure \ref{fig:TESS}.   ELC does not output light curves in the TESS bandpass so the $V$-band ELC light curve is roughly scaled to match the amplitude of the TESS light curve.  The models from ELC show that neither a main sequence nor a white dwarf companion will detectably eclipse the primary star (with the best-fit inclination of the system, $i = 69.3^\circ$).  The eclipse from the main-sequence companion would be indistinguishable from the impact of the starspots on the light curve unless a longer baseline of observations could be obtained and the signal detected over many rotations.

\begin{figure}
    \includegraphics[trim={2.5cm 5 2.5 5},clip,width=0.5\textwidth]{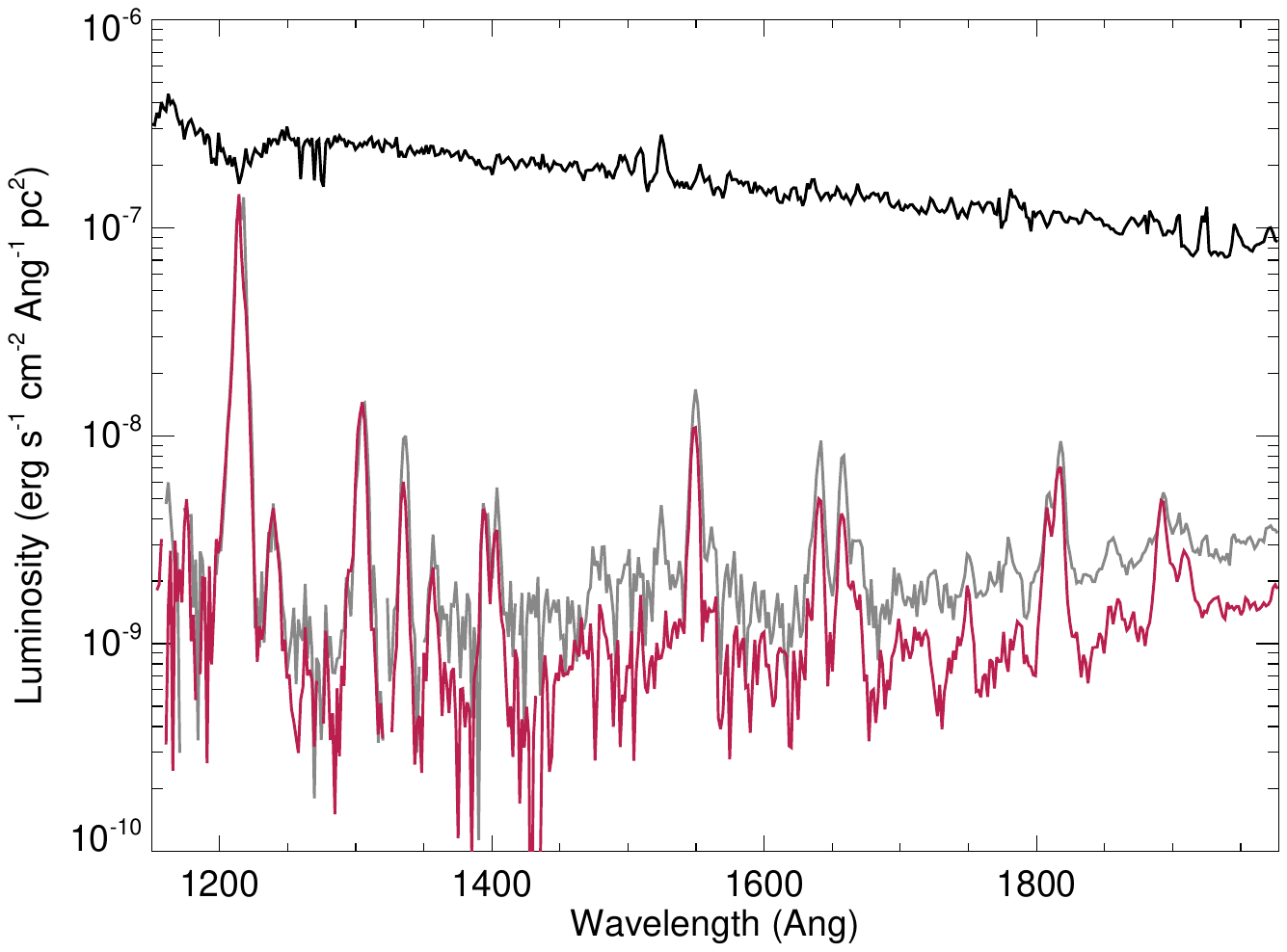}
    \caption{Observations from the Short-Wavelength Prime (SWP) spectrograph of IUE.  The black spectrum is an average SWP spectrum based on the data of 86 white dwarfs observed by IUE.  The gray spectrum is the average SWP spectrum based on the data of 34 RS CVn stars observed by IUE (excluding $\zeta$~And).  The red spectrum is the average SWP spectrum of $\zeta$~And. The features of the spectrum of $\zeta$~And are more closely aligned with those of the other RS CVns, including the large emission feature of Lyman-$\alpha$ ($1215.67$ \AA), which is seen in absorption for white dwarf stars.  
    } 
    \label{fig:IUE}
\end{figure}

We investigated the potential of an ultraviolet excess indicating the presence of a white dwarf companion using data from the International Ultraviolet Explorer \citep[IUE;][]{boggess1978}.    After accounting for stellar distance \citep{vanleeuwen07,gaia20,gaia23}, we found the average ultraviolet spectrum of the sample of 34 IUE target stars identified as RS CVns (not including $\zeta$~And) compiled into a catalog by \citet{beitiaantero2016}.  The ultraviolet flux and spectrum shape of observations from $\zeta$~And are not anomalous when compared to the average RS CVn spectrum (see Figure \ref{fig:IUE}).  When considering an average spectrum from the 86 targets identified as single white dwarfs in the same catalog, a white dwarf's ultraviolet flux amplitude and spectrum shape differ from that of $\zeta$~And.  From Figure \ref{fig:IUE}, it is clear that the IUE spectrum of $\zeta$~And is more consistent with that of an RS CVn and does not indicate the presence of a white dwarf.  

To further investigate the nature of the system, we considered the circularization and synchronization timescales.  From relations given by \citet{zahn1977}, the timescales are $t_\mathrm{circ} \sim 3 \times 10^{12}$ years and $t_\mathrm{sync} \sim 5 \times 10^9$ years, respectively.   
However, the system has an estimated age of $\sim0.5 \times 10^9$ years  \citep[][based on lithium abundance]{barradoynavascues1998} and is tidally locked \citep{fekel1999}, so circularization and synchronization have already occurred.  Because the age is less than the circularization and synchronization timescales, the system may have previously experienced mass transfer or mass loss.  
Due to the short mixing time for giant stars \citep{vlemmings2024}, the primary star is unlikely to show evidence of mass transfer with a high metallicity, consistent with $\zeta$~And having near-solar metallicity \citep[$\mathrm{[Fe/H]} = -0.05$;][]{luck2014}. 
Unfortunately, the stellar composition and tidally-locked orbit cannot further distinguish between a main-sequence and a white dwarf companion.

\section{Conclusions}

We have obtained long-baseline optical interferometric data from the CHARA Array for the spotted giant star $\zeta$~And.  Based on data spanning nearly six rotations of $\zeta$~And, we image the star at three epochs (approximately 1 rotation each).  We highlight similarities and differences from previous interferometric images of $\zeta$~And \citep{roettenbacher16b}.  While the large polar spot is still present, it is smaller and cooler than in previous images, but does appear to be growing during the duration of our observations.  These images also show the stellar surface evolving on a shorter timescale than previously expected in long-term photometry and Doppler imaging analyses.  
Based on the previous study by \citet{kovari2012}, we expected to be able to detect significant differential rotation over the course of our observations.  However,  starspot evolution occurs on a shorter timescale than is needed to detect differential rotation and may   account for the signal previously measured as differential rotation.  

We suggest that we cannot distinguish between the secondary star of $\zeta$~And being a main-sequence or a white dwarf companion.  The lack of an ultraviolet signature suggestive of a white dwarf is countered by our inability to detect the companion in our extensive interferometric data set.  Such detections were made in similar RS CVn systems with flux ratios between the primary and secondary stars comparable to what we expect for a main-sequence companion to $\zeta$~And \citep{roettenbacher15a,roettenbacher15b}.   While a main-sequence companion is more typical of RS CVn systems, a white dwarf companion is not a novel configuration; 29 Draconis (HD 160538), for example, was identified as an RS CVn having a white dwarf secondary by \citet{fekel1985}.  

In order to more completely study the evolution of the stellar surface of $\zeta$~And, higher cadence interferometric observations over a larger number of stellar rotations would be necessary.  A continuous data set would allow for the evolution of starspots to be tracked more carefully than the present data set, promoting a more detailed study of starspot evolution and potentially a clearer understanding of the star's differential rotation.  The suggested data set would be observationally expensive, but it would reveal unmatched insight for the stellar dynamos and rotation of active giant stars.

\begin{acknowledgments}
We thank the anonymous referee for their comments that improved the manuscript. The CHARA Array observing time was granted through the NOIRLab community-access program (NOIRLab Prop. ID: 2019A-0116; PI: R.\ Roettenbacher). The CHARA Array is supported by the National Science Foundation under Grant No. AST-1636624, AST-2034336, and AST-2407956, and the GSU College of Arts and Sciences, Office of the Provost, and Office of the Vice President for Research and Economic Development. 
MIRC-X received funding from the European Research Council (ERC) under the European Union’s Horizon 2020 research and innovation programme (Grant No. 639889). This research has made use of the Jean-Marie Mariotti Center \texttt{Aspro} service.  
This research was funded by the Hungarian National Research, Development and Innovation Office grant \'Elvonal KKP-143986.
This research has made use of the SIMBAD database, operated at CDS, Strasbourg, France.
The APT photometric data were supported by NASA, NSF, Tennessee State University, and the State of Tennessee through its Centers of Excellence program.  
RMR acknowledges support from the Heising-Simons Foundation's 51 Pegasi b Fellowship Program and NASA grants 80NSSC21K1034 and 80NSSC24K0108. JDM acknowledges funding for the development of MIRC-X (NASA-XRP NNX16AD43G, NSF-AST 1909165). S.K. acknowledges support from ERC Consolidator Grant (Grant Agreement ID 101003096) and STFC Small Award (ST/Y002695/1).

\end{acknowledgments}

\vspace{5mm}
\facilities{CHARA, TSU:APT, MAST (TESS, IUE), AAVSO, Konkoly}

\software{\texttt{astropy}, \texttt{astropy$\_$healpix}, \texttt{fitsh}, \texttt{healpy}, \texttt{mircx$\_$pipeline}, \texttt{SURFING}}

\bibliography{starspotpapers}{}

@ARTICLE{anugu20,
       author = {{Anugu}, Narsireddy and {Le Bouquin}, Jean-Baptiste and {Monnier}, John D. and {Kraus}, Stefan and {Setterholm}, Benjamin R. and {Labdon}, Aaron and {Davies}, Claire L. and {Lanthermann}, Cyprien and {Gardner}, Tyler and {Ennis}, Jacob and {Johnson}, Keith J.~C. and {Ten Brummelaar}, Theo and {Schaefer}, Gail and {Sturmann}, Judit},
        title = "{MIRC-X: A Highly Sensitive Six-telescope Interferometric Imager at the CHARA Array}",
      journal = {\aj},
         year = 2020,
        month = oct,
       volume = {160},
       number = {4},
          eid = {158},
        pages = {158},
          doi = {10.3847/1538-3881/aba957},
archivePrefix = {arXiv},
       eprint = {2007.12320},
 primaryClass = {astro-ph.IM},
       adsurl = {https://ui.adsabs.harvard.edu/abs/2020AJ....160..158A}
}

@ARTICLE{barradoynavascues1998,
       author = {{Barrado y Navascues}, D. and {de Castro}, E. and {Fernandez-Figueroa}, M.~J. and {Cornide}, M. and {Garcia Lopez}, R.~J.},
        title = "{The age-mass relation for chromospherically active binaries. III. Lithium depletion in giant components}",
      journal = {\aap},
         year = 1998,
        month = sep,
       volume = {337},
        pages = {739-753},
          doi = {10.48550/arXiv.astro-ph/9905244},
archivePrefix = {arXiv},
       eprint = {astro-ph/9905244},
 primaryClass = {astro-ph},
       adsurl = {https://ui.adsabs.harvard.edu/abs/1998A&A...337..739B}
}

@ARTICLE{beavers1986,
       author = {{Beavers}, W.~I. and {Eitter}, J.~J.},
        title = "{E. W. Fick Observatory Stellar Radial Velocity Measurements. I. 1976--1984}",
      journal = {\apjs},
         year = 1986,
        month = sep,
       volume = {62},
        pages = {147},
          doi = {10.1086/191136},
       adsurl = {https://ui.adsabs.harvard.edu/abs/1986ApJS...62..147B}
}

@ARTICLE{beitiaantero2016,
       author = {{Beitia-Antero}, Leire and {G{\'o}mez de Castro}, Ana I.},
        title = "{A database of synthetic photometry in the GALEX ultraviolet bands for the stellar sources observed with the International Ultraviolet Explorer}",
      journal = {\aap},
         year = 2016,
        month = nov,
       volume = {596},
          eid = {A49},
        pages = {A49},
          doi = {10.1051/0004-6361/201527782},
archivePrefix = {arXiv},
       eprint = {1605.04112},
 primaryClass = {astro-ph.IM},
       adsurl = {https://ui.adsabs.harvard.edu/abs/2016A&A...596A..49B}
}

@ARTICLE{boggess1978,
       author = {{Boggess}, A. and {Carr}, F.~A. and {Evans}, D.~C. and {Fischel}, D. and {Freeman}, H.~R. and {Fuechsel}, C.~F. and {Klinglesmith}, D.~A. and {Krueger}, V.~L. and {Longanecker}, G.~W. and {Moore}, J.~V.},
        title = "{The IUE spacecraft and instrumentation}",
      journal = {\nat},
         year = 1978,
        month = oct,
       volume = {275},
       number = {5679},
        pages = {372-377},
          doi = {10.1038/275372a0},
       adsurl = {https://ui.adsabs.harvard.edu/abs/1978Natur.275..372B}
}

@ARTICLE{campbell1911,
       author = {{Campbell}, William Wallace and {Moore}, Joseph Haines and {Wright}, William Hammond and {Duncan}, John Charles},
        title = "{Sixty-eight stars whose radial velocities vary}",
      journal = {Lick Observatory Bulletin},
         year = 1911,
        month = jan,
       volume = {199},
        pages = {140-154},
          doi = {10.5479/ADS/bib/1911LicOB.6.140C},
       adsurl = {https://ui.adsabs.harvard.edu/abs/1911LicOB...6..140C}
}

@ARTICLE{cannon1915,
       author = {{Cannon}, J.~B.},
        title = "{Orbit of {\ensuremath{\xi}} Andromed{\ae}}",
      journal = {\jrasc},
         year = 1915,
        month = apr,
       volume = {9},
        pages = {158},
       adsurl = {https://ui.adsabs.harvard.edu/abs/1915JRASC...9..158C}
}

@ARTICLE{chelli16,
       author = {{Chelli}, Alain and {Duvert}, Gilles and {Bourg{\`e}s}, Laurent and {Mella}, Guillaume and {Lafrasse}, Sylvain and {Bonneau}, Daniel and {Chesneau}, Olivier},
        title = "{Pseudomagnitudes and differential surface brightness: Application to the apparent diameter of stars}",
      journal = {\aap},
         year = 2016,
        month = may,
       volume = {589},
          eid = {A112},
        pages = {A112},
          doi = {10.1051/0004-6361/201527484},
archivePrefix = {arXiv},
       eprint = {1604.07700},
 primaryClass = {astro-ph.SR},
       adsurl = {https://ui.adsabs.harvard.edu/abs/2016A&A...589A.112C}
}

@ARTICLE{cutri03,
       author = {{Cutri}, R.~M. and {Skrutskie}, M.~F. and {van Dyk}, S. and {Beichman}, C.~A. and {Carpenter}, J.~M. and {Chester}, T. and {Cambresy}, L. and {Evans}, T. and {Fowler}, J. and {Gizis}, J. and {Howard}, E. and {Huchra}, J. and {Jarrett}, T. and {Kopan}, E.~L. and {Kirkpatrick}, J.~D. and {Light}, R.~M. and {Marsh}, K.~A. and {McCallon}, H. and {Schneider}, S. and {Stiening}, R. and {Sykes}, M. and {Weinberg}, M. and {Wheaton}, W.~A. and {Wheelock}, S. and {Zacarias}, N.},
        title = "{VizieR Online Data Catalog: 2MASS All-Sky Catalog of Point Sources (Cutri+ 2003)}",
      journal = {VizieR Online Data Catalog},
         year = 2003,
        month = jun,
          eid = {II/246},
        pages = {II/246},
       adsurl = {https://ui.adsabs.harvard.edu/abs/2003yCat.2246....0C}
}

@ARTICLE{ducati02,
       author = {{Ducati}, J.~R.},
        title = "{VizieR Online Data Catalog: Catalogue of Stellar Photometry in Johnson's 11-color system.}",
      journal = {VizieR Online Data Catalog},
         year = 2002,
        month = jan,
       adsurl = {https://ui.adsabs.harvard.edu/abs/2002yCat.2237....0D}
}

@ARTICLE{evans2024,
       author = {{Evans}, Nancy Remage and {Schaefer}, Gail H. and {Gallenne}, Alexandre and {Torres}, Guillermo and {Horch}, Elliott P. and {Anderson}, Richard I. and {Monnier}, John D. and {Roettenbacher}, Rachael M. and {Baron}, Fabien and {Anugu}, Narsireddy and {Davidson}, James W. and {Kervella}, Pierre and {Bras}, Garance and {Proffitt}, Charles and {M{\'e}rand}, Antoine and {Karovska}, Margarita and {Jones}, Jeremy and {Lanthermann}, Cyprien and {Kraus}, Stefan and {Codron}, Isabelle and {Bond}, Howard E. and {Viviani}, Giordano},
        title = "{The Orbit and Dynamical Mass of Polaris: Observations with the CHARA Array}",
      journal = {\apj},
         year = 2024,
        month = aug,
       volume = {971},
       number = {2},
          eid = {190},
        pages = {190},
          doi = {10.3847/1538-4357/ad5e7a},
archivePrefix = {arXiv},
       eprint = {2407.09641},
 primaryClass = {astro-ph.SR},
       adsurl = {https://ui.adsabs.harvard.edu/abs/2024ApJ...971..190E}
}

@ARTICLE{fekel1985,
       author = {{Fekel}, F.~C. and {Simon}, T.},
        title = "{HD 160538 and HD 185510: two active-chromosphere stars with hot companions.}",
      journal = {\aj},
         year = 1985,
        month = may,
       volume = {90},
        pages = {812-816},
          doi = {10.1086/113790},
       adsurl = {https://ui.adsabs.harvard.edu/abs/1985AJ.....90..812F}
}

@ARTICLE{fekel1999,
       author = {{Fekel}, F.~C. and {Strassmeier}, K.~G. and {Weber}, M. and {Washuettl}, A.},
        title = "{Orbital elements and physical parameters of ten chromospherically active binary stars}",
      journal = {\aaps},
         year = 1999,
        month = jun,
       volume = {137},
        pages = {369-383},
          doi = {10.1051/aas:1999252},
       adsurl = {https://ui.adsabs.harvard.edu/abs/1999A&AS..137..369F}
}

@ARTICLE{fekel2005,
       author = {{Fekel}, Francis C. and {Henry}, Gregory W. and {Lewis}, Ceteka},
        title = "{Chromospherically Active Stars. XXV. HD 144110=EV Draconis, a Double-lined Dwarf Binary}",
      journal = {\aj},
         year = 2005,
        month = aug,
       volume = {130},
       number = {2},
        pages = {794-798},
          doi = {10.1086/431316},
       adsurl = {https://ui.adsabs.harvard.edu/abs/2005AJ....130..794F}
}

@ARTICLE{gaia20,
       author = {{Gaia Collaboration} and {Brown}, A.~G.~A. and {Vallenari}, A. and {Prusti}, T. and {de Bruijne}, J.~H.~J. and {Babusiaux}, C. and {Biermann}, M.},
        title = "{Gaia Early Data Release 3: Summary of the contents and survey properties}",
      journal = {arXiv e-prints},
         year = 2020,
        month = dec,
          eid = {arXiv:2012.01533},
        pages = {arXiv:2012.01533},
archivePrefix = {arXiv},
       eprint = {2012.01533},
 primaryClass = {astro-ph.GA},
       adsurl = {https://ui.adsabs.harvard.edu/abs/2020arXiv201201533G}
}

@ARTICLE{gaia23,
       author = {{Gaia Collaboration} and {Vallenari}, A. and {Brown}, A.~G.~A. and {Prusti}, T. and {de Bruijne}, J.~H.~J. and {Arenou}, F. and {Babusiaux}, C. and {Biermann}, M. and {Creevey}, O.~L. and {Ducourant}, C. and {Evans}, D.~W. and {Eyer}, L. and {Guerra}, R. and {Hutton}, A. and {Jordi}, C. and {Klioner}, S.~A. and {Lammers}, U.~L. and {Lindegren}, L. and {Luri}, X. and {Mignard}, F. and {Panem}, C. and {Pourbaix}, D. and {Randich}, S. and {Sartoretti}, P. and {Soubiran}, C. and {Tanga}, P. and {Walton}, N.~A. and {Bailer-Jones}, C.~A.~L. and {Bastian}, U. and {Drimmel}, R. and {Jansen}, F. and {Katz}, D. and {Lattanzi}, M.~G. and {van Leeuwen}, F. and {Bakker}, J. and {Cacciari}, C. and {Casta{\~n}eda}, J. and {De Angeli}, F. and {Fabricius}, C. and {Fouesneau}, M. and {Fr{\'e}mat}, Y. and {Galluccio}, L. and {Guerrier}, A. and {Heiter}, U. and {Masana}, E. and {Messineo}, R. and {Mowlavi}, N. and {Nicolas}, C. and {Nienartowicz}, K. and {Pailler}, F. and {Panuzzo}, P. and {Riclet}, F. and {Roux}, W. and {Seabroke}, G.~M. and {Sordo}, R. and {Th{\'e}venin}, F. and {Gracia-Abril}, G. and {Portell}, J. and {Teyssier}, D. and {Altmann}, M. and {Andrae}, R. and {Audard}, M. and {Bellas-Velidis}, I. and {Benson}, K. and {Berthier}, J. and {Blomme}, R. and {Burgess}, P.~W. and {Busonero}, D. and {Busso}, G. and {C{\'a}novas}, H. and {Carry}, B. and {Cellino}, A. and {Cheek}, N. and {Clementini}, G. and {Damerdji}, Y. and {Davidson}, M. and {de Teodoro}, P. and {Nu{\~n}ez Campos}, M. and {Delchambre}, L. and {Dell'Oro}, A. and {Esquej}, P. and {Fern{\'a}ndez-Hern{\'a}ndez}, J. and {Fraile}, E. and {Garabato}, D. and {Garc{\'\i}a-Lario}, P. and {Gosset}, E. and {Haigron}, R. and {Halbwachs}, J. -L. and {Hambly}, N.~C. and {Harrison}, D.~L. and {Hern{\'a}ndez}, J. and {Hestroffer}, D. and {Hodgkin}, S.~T. and {Holl}, B. and {Jan{\ss}en}, K. and {Jevardat de Fombelle}, G. and {Jordan}, S. and {Krone-Martins}, A. and {Lanzafame}, A.~C. and {L{\"o}ffler}, W. and {Marchal}, O. and {Marrese}, P.~M. and {Moitinho}, A. and {Muinonen}, K. and {Osborne}, P. and {Pancino}, E. and {Pauwels}, T. and {Recio-Blanco}, A. and {Reyl{\'e}}, C. and {Riello}, M. and {Rimoldini}, L. and {Roegiers}, T. and {Rybizki}, J. and {Sarro}, L.~M. and {Siopis}, C. and {Smith}, M. and {Sozzetti}, A. and {Utrilla}, E. and {van Leeuwen}, M. and {Abbas}, U. and {{\'A}brah{\'a}m}, P. and {Abreu Aramburu}, A. and {Aerts}, C. and {Aguado}, J.~J. and {Ajaj}, M. and {Aldea-Montero}, F. and {Altavilla}, G. and {{\'A}lvarez}, M.~A. and {Alves}, J. and {Anders}, F. and {Anderson}, R.~I. and {Anglada Varela}, E. and {Antoja}, T. and {Baines}, D. and {Baker}, S.~G. and {Balaguer-N{\'u}{\~n}ez}, L. and {Balbinot}, E. and {Balog}, Z. and {Barache}, C. and {Barbato}, D. and {Barros}, M. and {Barstow}, M.~A. and {Bartolom{\'e}}, S. and {Bassilana}, J. -L. and {Bauchet}, N. and {Becciani}, U. and {Bellazzini}, M. and {Berihuete}, A. and {Bernet}, M. and {Bertone}, S. and {Bianchi}, L. and {Binnenfeld}, A. and {Blanco-Cuaresma}, S. and {Blazere}, A. and {Boch}, T. and {Bombrun}, A. and {Bossini}, D. and {Bouquillon}, S. and {Bragaglia}, A. and {Bramante}, L. and {Breedt}, E. and {Bressan}, A. and {Brouillet}, N. and {Brugaletta}, E. and {Bucciarelli}, B. and {Burlacu}, A. and {Butkevich}, A.~G. and {Buzzi}, R. and {Caffau}, E. and {Cancelliere}, R. and {Cantat-Gaudin}, T. and {Carballo}, R. and {Carlucci}, T. and {Carnerero}, M.~I. and {Carrasco}, J.~M. and {Casamiquela}, L. and {Castellani}, M. and {Castro-Ginard}, A. and {Chaoul}, L. and {Charlot}, P. and {Chemin}, L. and {Chiaramida}, V. and {Chiavassa}, A. and {Chornay}, N. and {Comoretto}, G. and {Contursi}, G. and {Cooper}, W.~J. and {Cornez}, T. and {Cowell}, S. and {Crifo}, F. and {Cropper}, M. and {Crosta}, M. and {Crowley}, C. and {Dafonte}, C. and {Dapergolas}, A. and {David}, M. and {David}, P. and {de Laverny}, P. and {De Luise}, F. and {De March}, R. and {De Ridder}, J. and {de Souza}, R. and {de Torres}, A. and {del Peloso}, E.~F. and {del Pozo}, E. and {Delbo}, M. and {Delgado}, A. and {Delisle}, J. -B. and {Demouchy}, C. and {Dharmawardena}, T.~E. and {Di Matteo}, P. and {Diakite}, S. and {Diener}, C. and {Distefano}, E. and {Dolding}, C. and {Edvardsson}, B. and {Enke}, H. and {Fabre}, C. and {Fabrizio}, M. and {Faigler}, S. and {Fedorets}, G. and {Fernique}, P. and {Fienga}, A. and {Figueras}, F. and {Fournier}, Y. and {Fouron}, C. and {Fragkoudi}, F. and {Gai}, M. and {Garcia-Gutierrez}, A. and {Garcia-Reinaldos}, M. and {Garc{\'\i}a-Torres}, M. and {Garofalo}, A. and {Gavel}, A. and {Gavras}, P. and {Gerlach}, E. and {Geyer}, R. and {Giacobbe}, P. and {Gilmore}, G. and {Girona}, S. and {Giuffrida}, G. and {Gomel}, R. and {Gomez}, A. and {Gonz{\'a}lez-N{\'u}{\~n}ez}, J. and {Gonz{\'a}lez-Santamar{\'\i}a}, I. and {Gonz{\'a}lez-Vidal}, J.~J. and {Granvik}, M. and {Guillout}, P. and {Guiraud}, J. and {Guti{\'e}rrez-S{\'a}nchez}, R. and {Guy}, L.~P. and {Hatzidimitriou}, D. and {Hauser}, M. and {Haywood}, M. and {Helmer}, A. and {Helmi}, A. and {Sarmiento}, M.~H. and {Hidalgo}, S.~L. and {Hilger}, T. and {H{\l}adczuk}, N. and {Hobbs}, D. and {Holland}, G. and {Huckle}, H.~E. and {Jardine}, K. and {Jasniewicz}, G. and {Jean-Antoine Piccolo}, A. and {Jim{\'e}nez-Arranz}, {\'O}. and {Jorissen}, A. and {Juaristi Campillo}, J. and {Julbe}, F. and {Karbevska}, L. and {Kervella}, P. and {Khanna}, S. and {Kontizas}, M. and {Kordopatis}, G. and {Korn}, A.~J. and {K{\'o}sp{\'a}l}, {\'A}. and {Kostrzewa-Rutkowska}, Z. and {Kruszy{\'n}ska}, K. and {Kun}, M. and {Laizeau}, P. and {Lambert}, S. and {Lanza}, A.~F. and {Lasne}, Y. and {Le Campion}, J. -F. and {Lebreton}, Y. and {Lebzelter}, T. and {Leccia}, S. and {Leclerc}, N. and {Lecoeur-Taibi}, I. and {Liao}, S. and {Licata}, E.~L. and {Lindstr{\o}m}, H.~E.~P. and {Lister}, T.~A. and {Livanou}, E. and {Lobel}, A. and {Lorca}, A. and {Loup}, C. and {Madrero Pardo}, P. and {Magdaleno Romeo}, A. and {Managau}, S. and {Mann}, R.~G. and {Manteiga}, M. and {Marchant}, J.~M. and {Marconi}, M. and {Marcos}, J. and {Marcos Santos}, M.~M.~S. and {Mar{\'\i}n Pina}, D. and {Marinoni}, S. and {Marocco}, F. and {Marshall}, D.~J. and {Martin Polo}, L. and {Mart{\'\i}n-Fleitas}, J.~M. and {Marton}, G. and {Mary}, N. and {Masip}, A. and {Massari}, D. and {Mastrobuono-Battisti}, A. and {Mazeh}, T. and {McMillan}, P.~J. and {Messina}, S. and {Michalik}, D. and {Millar}, N.~R. and {Mints}, A. and {Molina}, D. and {Molinaro}, R. and {Moln{\'a}r}, L. and {Monari}, G. and {Mongui{\'o}}, M. and {Montegriffo}, P. and {Montero}, A. and {Mor}, R. and {Mora}, A. and {Morbidelli}, R. and {Morel}, T. and {Morris}, D. and {Muraveva}, T. and {Murphy}, C.~P. and {Musella}, I. and {Nagy}, Z. and {Noval}, L. and {Oca{\~n}a}, F. and {Ogden}, A. and {Ordenovic}, C. and {Osinde}, J.~O. and {Pagani}, C. and {Pagano}, I. and {Palaversa}, L. and {Palicio}, P.~A. and {Pallas-Quintela}, L. and {Panahi}, A. and {Payne-Wardenaar}, S. and {Pe{\~n}alosa Esteller}, X. and {Penttil{\"a}}, A. and {Pichon}, B. and {Piersimoni}, A.~M. and {Pineau}, F. -X. and {Plachy}, E. and {Plum}, G. and {Poggio}, E. and {Pr{\v{s}}a}, A. and {Pulone}, L. and {Racero}, E. and {Ragaini}, S. and {Rainer}, M. and {Raiteri}, C.~M. and {Rambaux}, N. and {Ramos}, P. and {Ramos-Lerate}, M. and {Re Fiorentin}, P. and {Regibo}, S. and {Richards}, P.~J. and {Rios Diaz}, C. and {Ripepi}, V. and {Riva}, A. and {Rix}, H. -W. and {Rixon}, G. and {Robichon}, N. and {Robin}, A.~C. and {Robin}, C. and {Roelens}, M. and {Rogues}, H.~R.~O. and {Rohrbasser}, L. and {Romero-G{\'o}mez}, M. and {Rowell}, N. and {Royer}, F. and {Ruz Mieres}, D. and {Rybicki}, K.~A. and {Sadowski}, G. and {S{\'a}ez N{\'u}{\~n}ez}, A. and {Sagrist{\`a} Sell{\'e}s}, A. and {Sahlmann}, J. and {Salguero}, E. and {Samaras}, N. and {Sanchez Gimenez}, V. and {Sanna}, N. and {Santove{\~n}a}, R. and {Sarasso}, M. and {Schultheis}, M. and {Sciacca}, E. and {Segol}, M. and {Segovia}, J.~C. and {S{\'e}gransan}, D. and {Semeux}, D. and {Shahaf}, S. and {Siddiqui}, H.~I. and {Siebert}, A. and {Siltala}, L. and {Silvelo}, A. and {Slezak}, E. and {Slezak}, I. and {Smart}, R.~L. and {Snaith}, O.~N. and {Solano}, E. and {Solitro}, F. and {Souami}, D. and {Souchay}, J. and {Spagna}, A. and {Spina}, L. and {Spoto}, F. and {Steele}, I.~A. and {Steidelm{\"u}ller}, H. and {Stephenson}, C.~A. and {S{\"u}veges}, M. and {Surdej}, J. and {Szabados}, L. and {Szegedi-Elek}, E. and {Taris}, F. and {Taylor}, M.~B. and {Teixeira}, R. and {Tolomei}, L. and {Tonello}, N. and {Torra}, F. and {Torra}, J. and {Torralba Elipe}, G. and {Trabucchi}, M. and {Tsounis}, A.~T. and {Turon}, C. and {Ulla}, A. and {Unger}, N. and {Vaillant}, M.~V. and {van Dillen}, E. and {van Reeven}, W. and {Vanel}, O. and {Vecchiato}, A. and {Viala}, Y. and {Vicente}, D. and {Voutsinas}, S. and {Weiler}, M. and {Wevers}, T. and {Wyrzykowski}, {\L}. and {Yoldas}, A. and {Yvard}, P. and {Zhao}, H. and {Zorec}, J. and {Zucker}, S. and {Zwitter}, T.},
        title = "{Gaia Data Release 3. Summary of the content and survey properties}",
      journal = {\aap},
         year = 2023,
        month = jun,
       volume = {674},
          eid = {A1},
        pages = {A1},
          doi = {10.1051/0004-6361/202243940},
archivePrefix = {arXiv},
       eprint = {2208.00211},
 primaryClass = {astro-ph.GA},
       adsurl = {https://ui.adsabs.harvard.edu/abs/2023A&A...674A...1G}
}

@ARTICLE{gardner22,
       author = {{Gardner}, Tyler and {Monnier}, John D. and {Fekel}, Francis C. and {Le Bouquin}, Jean-Baptiste and {Scovera}, Adam and {Schaefer}, Gail and {Kraus}, Stefan and {Adams}, Fred C. and {Anugu}, Narsireddy and {Berger}, Jean-Philippe and {Ten Brummelaar}, Theo and {Davies}, Claire L. and {Ennis}, Jacob and {Gies}, Douglas R. and {Johnson}, Keith J.~C. and {Kervella}, Pierre and {Kratter}, Kaitlin M. and {Labdon}, Aaron and {Lanthermann}, Cyprien and {Sahlmann}, Johannes and {Setterholm}, Benjamin R.},
        title = "{ARMADA. II. Further Detections of Inner Companions to Intermediate-mass Binaries with Microarcsecond Astrometry at CHARA and VLTI}",
      journal = {\aj},
         year = 2022,
        month = nov,
       volume = {164},
       number = {5},
          eid = {184},
        pages = {184},
          doi = {10.3847/1538-3881/ac8eae},
archivePrefix = {arXiv},
       eprint = {2209.00669},
 primaryClass = {astro-ph.SR},
       adsurl = {https://ui.adsabs.harvard.edu/abs/2022AJ....164..184G}
}

@ARTICLE{gratton1950,
       author = {{Gratton}, Livio},
        title = "{C{\ensuremath{\alpha}} II Emission in Lambda and Zeta Andromedae.}",
      journal = {\apj},
         year = 1950,
        month = jan,
       volume = {111},
        pages = {31},
          doi = {10.1086/145236},
       adsurl = {https://ui.adsabs.harvard.edu/abs/1950ApJ...111...31G}
}

@INPROCEEDINGS{hall1976,
       author = {{Hall}, D.~S.},
        title = "{The RS CVn Binaries and Binaries with Similar PROPERTI{\'E}S}",
    booktitle = {IAU Colloq. 29: Multiple Periodic Variable Stars},
         year = 1976,
       editor = {{Fitch}, Walter S.},
       series = {Astrophysics and Space Science Library},
       volume = {60},
        month = jan,
        pages = {287},
          doi = {10.1007/978-94-010-1175-4_15},
       adsurl = {https://ui.adsabs.harvard.edu/abs/1976ASSL...60..287H}
}

@ARTICLE{harmon00,
       author = {{Harmon}, Robert O. and {Crews}, Lionel J.},
        title = "{Imaging Stellar Surfaces via Matrix Light-Curve Inversion}",
      journal = {\aj},
         year = 2000,
        month = dec,
       volume = {120},
       number = {6},
        pages = {3274-3294},
          doi = {10.1086/316882},
       adsurl = {https://ui.adsabs.harvard.edu/abs/2000AJ....120.3274H}
}

@ARTICLE{hathaway2015,
       author = {{Hathaway}, David H.},
        title = "{The Solar Cycle}",
      journal = {Living Reviews in Solar Physics},
         year = 2015,
        month = dec,
       volume = {12},
       number = {1},
          eid = {4},
        pages = {4},
          doi = {10.1007/lrsp-2015-4},
archivePrefix = {arXiv},
       eprint = {1502.07020},
 primaryClass = {astro-ph.SR},
       adsurl = {https://ui.adsabs.harvard.edu/abs/2015LRSP...12....4H}
}

@ARTICLE{hendry1980,
       author = {{Hendry}, E.~M.},
        title = "{Recent changes in the CA II H and K emission lines in the spectrum of ZET And.}",
      journal = {\pasp},
         year = 1980,
        month = dec,
       volume = {92},
        pages = {825-828},
          doi = {10.1086/130757},
       adsurl = {https://ui.adsabs.harvard.edu/abs/1980PASP...92..825H}
}

@ARTICLE{henry1995,
       author = {{Henry}, Gregory W. and {Eaton}, Joel A. and {Hamer}, Jamesia and
         {Hall}, Douglas S.},
        title = "{Starspot Evolution, Differential Rotation, and Magnetic Cycles in the Chromospherically Active Binaries lambda Andromedae, sigma Geminorum, II Pegasi, and V711 Tauri}",
      journal = {\apjs},
         year = "1995",
        month = "Apr",
       volume = {97},
        pages = {513},
          doi = {10.1086/192149},
       adsurl = {https://ui.adsabs.harvard.edu/abs/1995ApJS...97..513H}
}

@ARTICLE{henry1999,
       author = {{Henry}, Gregory W.},
        title = "{Techniques for Automated High-Precision Photometry of Sun-like Stars}",
      journal = {\pasp},
         year = 1999,
        month = jul,
       volume = {111},
       number = {761},
        pages = {845-860},
          doi = {10.1086/316388},
       adsurl = {https://ui.adsabs.harvard.edu/abs/1999PASP..111..845H}
}

@ARTICLE{kaye95,
       author = {{Kaye}, A.~B. and {Hall}, D.~S. and {Henry}, G.~W. and {Eaton}, J.~A. and {Lines}, R.~D. and {Lines}, H.~C. and {Barksdale}, W.~S. and {Beck}, S.~J. and {Chambliss}, C.~R. and {Fried}, R.~E. and {Genet}, R.~M. and {Hopkins}, J.~L. and {Lovell}, L.~P. and {Louth}, H.~P. and {Montle}, R.~E. and {Renner}, T.~R. and {Stelzer}, H.~J.},
        title = "{Suspected Starspots Found on the K Giants in Seven Ellipsoidal RS CVn-Type Binaries}",
      journal = {\aj},
         year = 1995,
        month = may,
       volume = {109},
        pages = {2177},
          doi = {10.1086/117443},
       adsurl = {https://ui.adsabs.harvard.edu/abs/1995AJ....109.2177K}
}

@unpublished{kloppenborg25,
    author = {{Kloppenborg},Brian} ,
    title = {Observations from the AAVSO International Database},
    year = 2025,
    url = {https://www.aavso.org}
}

@ARTICLE{korhonen10,
       author = {{Korhonen}, H. and {Wittkowski}, M. and {Kov{\'a}ri}, Zs. and
         {Granzer}, Th. and {Hackman}, T. and {Strassmeier}, K.~G.},
        title = "{Ellipsoidal primary of the RS CVn binary {\ensuremath{\zeta}} Andromedae . Investigation using high-resolution spectroscopy and optical interferometry}",
      journal = {\aap},
         year = 2010,
        month = jun,
       volume = {515},
          eid = {A14},
        pages = {A14},
          doi = {10.1051/0004-6361/200913736},
archivePrefix = {arXiv},
       eprint = {1002.4201},
 primaryClass = {astro-ph.SR},
       adsurl = {https://ui.adsabs.harvard.edu/abs/2010A&A...515A..14K}
}

@ARTICLE{kovari07,
       author = {{K\H{o}v{\'a}ri}, {\protect Zs}. and {Bartus}, J. and {{\v{S}}vanda}, M. and
         {Vida}, K. and {Strassmeier}, K.~G. and {Ol{\'a}h}, K. and
         {Forg{\'a}cs-Dajka}, E.},
        title = "{Surface velocity network with anti-solar differential rotation on the active K-giant {\ensuremath{\sigma}} Geminorum}",
      journal = {Astronomische Nachrichten},
         year = 2007,
        month = dec,
       volume = {328},
       number = {10},
        pages = {1081},
          doi = {10.1002/asna.200710874},
archivePrefix = {arXiv},
       eprint = {0711.0771},
 primaryClass = {astro-ph},
       adsurl = {https://ui.adsabs.harvard.edu/abs/2007AN....328.1081K}
}

@ARTICLE{kovari2012,
       author = {{K{\H{o}}v{\'a}ri}, {\protect Zs} and {Korhonen}, H. and {Kriskovics}, L. and {Vida}, K. and {Donati}, J. -F. and {Le Coroller}, H. and {Monnier}, J.~D. and {Pedretti}, E. and {Petit}, P.},
        title = "{Measuring differential rotation of the K-giant {\ensuremath{\zeta}} Andromedae}",
      journal = {\aap},
         year = 2012,
        month = mar,
       volume = {539},
          eid = {A50},
        pages = {A50},
          doi = {10.1051/0004-6361/201118177},
archivePrefix = {arXiv},
       eprint = {1201.2921},
 primaryClass = {astro-ph.SR},
       adsurl = {https://ui.adsabs.harvard.edu/abs/2012A&A...539A..50K}
}

@ARTICLE{kron47,
       author = {{Kron}, Gerald E.},
        title = "{The Probable Detecting of Surface Spots on AR Lacertae B}",
      journal = {\pasp},
         year = 1947,
        month = oct,
       volume = {59},
       number = {350},
        pages = {261},
          doi = {10.1086/125964},
       adsurl = {https://ui.adsabs.harvard.edu/abs/1947PASP...59..261K}
}

@ARTICLE{kron50,
       author = {{Kron}, G.~E.},
        title = "{Special characteristics of a few late-type dwarf stars.}",
      journal = {\aj},
         year = 1950,
        month = apr,
       volume = {55},
          eid = {69-69},
        pages = {69-69},
       adsurl = {https://ui.adsabs.harvard.edu/abs/1950AJ.....55...69K}
}

@ARTICLE{luck2014,
       author = {{Luck}, R. Earle},
        title = "{Parameters and Abundances in Luminous Stars}",
      journal = {\aj},
         year = 2014,
        month = jun,
       volume = {147},
       number = {6},
          eid = {137},
        pages = {137},
          doi = {10.1088/0004-6256/147/6/137},
       adsurl = {https://ui.adsabs.harvard.edu/abs/2014AJ....147..137L}
}

@ARTICLE{martinez21,
       author = {{Martinez}, Arturo O. and {Baron}, Fabien R. and {Monnier}, John D. and {Roettenbacher}, Rachael M. and {Parks}, J. Robert},
        title = "{Dynamical Surface Imaging of {\ensuremath{\lambda}} Andromedae}",
      journal = {\apj},
         year = 2021,
        month = jul,
       volume = {916},
       number = {1},
          eid = {60},
        pages = {60},
          doi = {10.3847/1538-4357/ac06a5},
archivePrefix = {arXiv},
       eprint = {2107.06366},
 primaryClass = {astro-ph.SR},
       adsurl = {https://ui.adsabs.harvard.edu/abs/2021ApJ...916...60M}
}

@ARTICLE{massarotti2008,
       author = {{Massarotti}, Alessandro and {Latham}, David W. and {Stefanik}, Robert P. and {Fogel}, Jeffrey},
        title = "{Rotational and Radial Velocities for a Sample of 761 HIPPARCOS Giants and the Role of Binarity}",
      journal = {\aj},
         year = 2008,
        month = jan,
       volume = {135},
       number = {1},
        pages = {209-231},
          doi = {10.1088/0004-6256/135/1/209},
       adsurl = {https://ui.adsabs.harvard.edu/abs/2008AJ....135..209M}
}

@INPROCEEDINGS{monnier04,
       author = {{Monnier}, John D. and {Berger}, Jean-Philippe and {Millan-Gabet}, Rafael and {ten Brummelaar}, Theo A.},
        title = "{The Michigan Infrared Combiner (MIRC): IR imaging with the CHARA Array}",
    booktitle = {New Frontiers in Stellar Interferometry},
         year = 2004,
       editor = {{Traub}, Wesley A.},
       series = {Society of Photo-Optical Instrumentation Engineers (SPIE) Conference Series},
       volume = {5491},
        month = oct,
        pages = {1370},
          doi = {10.1117/12.550804},
       adsurl = {https://ui.adsabs.harvard.edu/abs/2004SPIE.5491.1370M}
}

@ARTICLE{monnier07,
       author = {{Monnier}, John D. and {Zhao}, M. and {Pedretti}, E. and {Thureau}, N. and
         {Ireland}, M. and {Muirhead}, P. and {Berger}, J. -P. and
         {Millan-Gabet}, R. and {Van Belle}, G. and {ten Brummelaar}, T. and
         {McAlister}, H. and {Ridgway}, S. and {Turner}, N. and {Sturmann}, L. and
         {Sturmann}, J. and {Berger}, D.},
        title = "{Imaging the Surface of Altair}",
      journal = {Science},
         year = 2007,
        month = jul,
       volume = {317},
       number = {5836},
        pages = {342},
          doi = {10.1126/science.1143205},
archivePrefix = {arXiv},
       eprint = {0706.0867},
 primaryClass = {astro-ph},
       adsurl = {https://ui.adsabs.harvard.edu/abs/2007Sci...317..342M}
}

@ARTICLE{monnier12,
       author = {{Monnier}, J.~D. and {Che}, Xiao and {Zhao}, Ming and {Ekstr{\"o}m}, S. and {Maestro}, V. and {Aufdenberg}, Jason and {Baron}, F. and {Georgy}, C. and {Kraus}, S. and {McAlister}, H. and {Pedretti}, E. and {Ridgway}, S. and {Sturmann}, J. and {Sturmann}, L. and {ten Brummelaar}, T. and {Thureau}, N. and {Turner}, N. and {Tuthill}, P.~G.},
        title = "{Resolving Vega and the Inclination Controversy with CHARA/MIRC}",
      journal = {\apjl},
         year = 2012,
        month = dec,
       volume = {761},
       number = {1},
          eid = {L3},
        pages = {L3},
          doi = {10.1088/2041-8205/761/1/L3},
archivePrefix = {arXiv},
       eprint = {1211.6055},
 primaryClass = {astro-ph.SR},
       adsurl = {https://ui.adsabs.harvard.edu/abs/2012ApJ...761L...3M}
}

@ARTICLE{orosz2000,
       author = {{Orosz}, J.~A. and {Hauschildt}, P.~H.},
        title = "{The use of the NextGen model atmospheres for cool giants in a light curve synthesis code}",
      journal = {\aap},
         year = 2000,
        month = dec,
       volume = {364},
        pages = {265-281},
          doi = {10.48550/arXiv.astro-ph/0010114},
archivePrefix = {arXiv},
       eprint = {astro-ph/0010114},
 primaryClass = {astro-ph},
       adsurl = {https://ui.adsabs.harvard.edu/abs/2000A&A...364..265O}
}

@ARTICLE{rice89,
       author = {{Rice}, J.~B. and {Wehlau}, W.~H. and {Khokhlova}, V.~L.},
        title = "{Mapping stellar surfaces by Doppler imaging : technique and application.}",
      journal = {\aap},
         year = "1989",
        month = "Jan",
       volume = {208},
        pages = {179-188},
       adsurl = {https://ui.adsabs.harvard.edu/abs/1989A&A...208..179R}
}

@INPROCEEDINGS{ricker14,
       author = {{Ricker}, George R. and {Winn}, Joshua N. and {Vanderspek}, Roland and
         {Latham}, David W. and {Bakos}, G{\'a}sp{\'a}r. {\'A}. and
         {Bean}, Jacob L. and {Berta-Thompson}, Zachory K. and
         {Brown}, Timothy M. and {Buchhave}, Lars and {Butler}, Nathaniel R. and
         {Butler}, R. Paul and {Chaplin}, William J. and {Charbonneau}, David and
         {Christensen-Dalsgaard}, J{\o}rgen and {Clampin}, Mark and
         {Deming}, Drake and {Doty}, John and {De Lee}, Nathan and
         {Dressing}, Courtney and {Dunham}, E.~W. and {Endl}, Michael and
         {Fressin}, Francois and {Ge}, Jian and {Henning}, Thomas and
         {Holman}, Matthew J. and {Howard}, Andrew W. and {Ida}, Shigeru and
         {Jenkins}, Jon and {Jernigan}, Garrett and {Johnson}, John A. and
         {Kaltenegger}, Lisa and {Kawai}, Nobuyuki and {Kjeldsen}, Hans and
         {Laughlin}, Gregory and {Levine}, Alan M. and {Lin}, Douglas and
         {Lissauer}, Jack J. and {MacQueen}, Phillip and {Marcy}, Geoffrey and
         {McCullough}, P.~R. and {Morton}, Timothy D. and {Narita}, Norio and
         {Paegert}, Martin and {Palle}, Enric and {Pepe}, Francesco and
         {Pepper}, Joshua and {Quirrenbach}, Andreas and {Rinehart}, S.~A. and
         {Sasselov}, Dimitar and {Sato}, Bun'ei and {Seager}, Sara and
         {Sozzetti}, Alessandro and {Stassun}, Keivan G. and {Sullivan}, Peter and
         {Szentgyorgyi}, Andrew and {Torres}, Guillermo and {Udry}, Stephane and
         {Villasenor}, Joel},
        title = "{Transiting Exoplanet Survey Satellite (TESS)}",
    booktitle = {\procspie},
         year = 2014,
       series = {Society of Photo-Optical Instrumentation Engineers (SPIE) Conference Series},
       volume = {9143},
        month = aug,
          eid = {914320},
        pages = {914320},
          doi = {10.1117/12.2063489},
archivePrefix = {arXiv},
       eprint = {1406.0151},
 primaryClass = {astro-ph.EP},
       adsurl = {https://ui.adsabs.harvard.edu/abs/2014SPIE.9143E..20R}
}

@ARTICLE{roettenbacher11,
       author = {{Roettenbacher}, Rachael M. and {Harmon}, Robert O. and
         {Vutisalchavakul}, Nalin and {Henry}, Gregory W.},
        title = "{A Study of Differential Rotation on II Pegasi via Photometric Starspot Imaging}",
      journal = {\aj},
         year = 2011,
        month = apr,
       volume = {141},
       number = {4},
          eid = {138},
        pages = {138},
          doi = {10.1088/0004-6256/141/4/138},
archivePrefix = {arXiv},
       eprint = {1009.2308},
 primaryClass = {astro-ph.SR},
       adsurl = {https://ui.adsabs.harvard.edu/abs/2011AJ....141..138R}
}

@ARTICLE{roettenbacher15a,
       author = {{Roettenbacher}, Rachael M. and {Monnier}, John D. and {Fekel}, Francis C. and {Henry}, Gregory W. and {Korhonen}, Heidi and {Latham}, David W. and {Muterspaugh}, Matthew W. and {Williamson}, Michael H. and {Baron}, Fabien and {ten Brummelaar}, Theo A. and {Che}, Xiao and {Harmon}, Robert O. and {Schaefer}, Gail H. and {Scott}, Nicholas J. and {Sturmann}, Judit and {Sturmann}, Laszlo and {Turner}, Nils H.},
        title = "{Detecting the Companions and Ellipsoidal Variations of RS CVn Primaries. II. o Draconis, a Candidate for Recent Low-mass Companion Ingestion}",
      journal = {\apj},
         year = 2015,
        month = aug,
       volume = {809},
       number = {2},
          eid = {159},
        pages = {159},
          doi = {10.1088/0004-637X/809/2/159},
archivePrefix = {arXiv},
       eprint = {1507.03601},
 primaryClass = {astro-ph.SR},
       adsurl = {https://ui.adsabs.harvard.edu/abs/2015ApJ...809..159R}
}

@ARTICLE{roettenbacher15b,
       author = {{Roettenbacher}, Rachael M. and {Monnier}, John D. and {Henry}, Gregory W. and {Fekel}, Francis C. and {Williamson}, Michael H. and {Pourbaix}, Dimitri and {Latham}, David W. and {Latham}, Christian A. and {Torres}, Guillermo and {Baron}, Fabien and {Che}, Xiao and {Kraus}, Stefan and {Schaefer}, Gail H. and {Aarnio}, Alicia N. and {Korhonen}, Heidi and {Harmon}, Robert O. and {ten Brummelaar}, Theo A. and {Sturmann}, Judit and {Sturmann}, Laszlo and {Turner}, Nils H.},
        title = "{Detecting the Companions and Ellipsoidal Variations of RS CVn Primaries. I. {\ensuremath{\sigma}} Geminorum}",
      journal = {\apj},
         year = 2015,
        month = jul,
       volume = {807},
       number = {1},
          eid = {23},
        pages = {23},
          doi = {10.1088/0004-637X/807/1/23},
archivePrefix = {arXiv},
       eprint = {1504.06628},
 primaryClass = {astro-ph.SR},
       adsurl = {https://ui.adsabs.harvard.edu/abs/2015ApJ...807...23R}
}

@ARTICLE{roettenbacher16b,
   author = {{Roettenbacher}, R.~M. and {Monnier}, J.~D. and {Korhonen}, H. and 
	{Aarnio}, A.~N. and {Baron}, F. and {Che}, X. and {Harmon}, R.~O. and 
	{K{\H o}v{\'a}ri}, Z. and {Kraus}, S. and {Schaefer}, G.~H. and 
	{Torres}, G. and {Zhao}, M. and {Ten Brummelaar}, T.~A. and 
	{Sturmann}, J. and {Sturmann}, L.},
    title = "{No Sun-like dynamo on the active star {$\zeta$} Andromedae from starspot asymmetry}",
  journal = {\nat},
archivePrefix = "arXiv",
   eprint = {1709.10107},
 primaryClass = "astro-ph.SR",
     year = 2016,
    month = may,
   volume = 533,
    pages = {217-220},
      doi = {10.1038/nature17444},
   adsurl = {http://adsabs.harvard.edu/abs/2016Natur.533..217R}
}

@ARTICLE{roettenbacher17b,
   author = {{Roettenbacher}, R.~M. and {Monnier}, J.~D. and {Korhonen}, H. and 
	{Harmon}, R.~O. and {Baron}, F. and {Hackman}, T. and {Henry}, G.~W. and 
	{Schaefer}, G.~H. and {Strassmeier}, K.~G. and {Weber}, M. and 
	{ten Brummelaar}, T.~A.},
    title = "{Contemporaneous Imaging Comparisons of the Spotted Giant {$\sigma$} Geminorum Using Interferometric, Spectroscopic, and Photometric Data}",
  journal = {\apj},
archivePrefix = "arXiv",
   eprint = {1709.10109},
 primaryClass = "astro-ph.SR",
     year = 2017,
    month = nov,
   volume = 849,
      eid = {120},
    pages = {120},
      doi = {10.3847/1538-4357/aa8ef7},
   adsurl = {http://adsabs.harvard.edu/abs/2017ApJ...849..120R}
}

@ARTICLE{SpencerJones1928,
       author = {{Spencer Jones}, Harold},
        title = "{Radial velocity determinations : including a spectroscopic determination of the constant of aberration, the orbits of 13 spectroscopic binary stars, and the radial velocities of 434 stars}",
      journal = {Annals of the Cape Observatory},
         year = 1928,
        month = jan,
       volume = {10},
        pages = {8},
       adsurl = {https://ui.adsabs.harvard.edu/abs/1928AnCap..10....8S}
}

@ARTICLE{stawikowski1994,
       author = {{Stawikowski}, A. and {Glebocki}, R.},
        title = "{Are the Rotational Axes Perpendicular to the Orbital Planes in the Binary Systems? II. Synchronous Long-Period RS CVn Stars}",
      journal = {\actaa},
         year = 1994,
        month = oct,
       volume = {44},
        pages = {393-406},
       adsurl = {https://ui.adsabs.harvard.edu/abs/1994AcA....44..393S}
}

@ARTICLE{stebbins1928,
       author = {{Stebbins}, Joel},
        title = "{Photo-Electric Photometry of Stars. Chapter IV. {\ensuremath{\zeta}} Andromedae.}",
      journal = {Publications of the Washburn Observatory},
         year = 1928,
        month = jan,
       volume = {15},
        pages = {29-32},
       adsurl = {https://ui.adsabs.harvard.edu/abs/1928PWasO..15...29S}
}

@ARTICLE{strassmeier89,
       author = {{Strassmeier}, Klaus G. and {Hall}, Douglas S. and {Boyd}, Louis J. and {Genet}, Russell M.},
        title = "{Photometric Variability in Chromospherically Active Stars. III. The Binary Stars}",
      journal = {\apjs},
         year = 1989,
        month = jan,
       volume = {69},
        pages = {141},
          doi = {10.1086/191310},
       adsurl = {https://ui.adsabs.harvard.edu/abs/1989ApJS...69..141S}
}

@ARTICLE{strassmeier1999,
       author = {{Strassmeier}, K.~G.},
        title = "{Doppler imaging of stellar surface structure. XI. The super starspots on the K0 giant HD 12545: larger than the entire Sun}",
      journal = {\aap},
         year = 1999,
        month = jul,
       volume = {347},
        pages = {225-234},
       adsurl = {https://ui.adsabs.harvard.edu/abs/1999A&A...347..225S}
}

@ARTICLE{strassmeier1999c,
       author = {{Strassmeier}, K.~G. and {Serkowitsch}, E. and {Granzer}, Th.},
        title = "{Starspot photometry with robotic telescopes. UBV(RI)\_C and by light curves of 47 active stars in 1996/97}",
      journal = {\aaps},
         year = 1999,
        month = nov,
       volume = {140},
        pages = {29-53},
          doi = {10.1051/aas:1999116},
       adsurl = {https://ui.adsabs.harvard.edu/abs/1999A&AS..140...29S}
}

@ARTICLE{tenbrummelaar05,
       author = {{ten Brummelaar}, T.~A. and {McAlister}, H.~A. and {Ridgway}, S.~T. and {Bagnuolo}, W.~G., Jr. and {Turner}, N.~H. and {Sturmann}, L. and {Sturmann}, J. and {Berger}, D.~H. and {Ogden}, C.~E. and {Cadman}, R. and {Hartkopf}, W.~I. and {Hopper}, C.~H. and {Shure}, M.~A.},
        title = "{First Results from the CHARA Array. II. A Description of the Instrument}",
      journal = {\apj},
         year = 2005,
        month = jul,
       volume = {628},
       number = {1},
        pages = {453-465},
          doi = {10.1086/430729},
archivePrefix = {arXiv},
       eprint = {astro-ph/0504082},
 primaryClass = {astro-ph},
       adsurl = {https://ui.adsabs.harvard.edu/abs/2005ApJ...628..453T}
}

@ARTICLE{vanhamme1993,
       author = {{van Hamme}, W.},
        title = "{New Limb-Darkening Coefficients for Modeling Binary Star Light Curves}",
      journal = {\aj},
         year = 1993,
        month = nov,
       volume = {106},
        pages = {2096},
          doi = {10.1086/116788},
       adsurl = {https://ui.adsabs.harvard.edu/abs/1993AJ....106.2096V}
}

@ARTICLE{vanleeuwen07,
       author = {{van Leeuwen}, F.},
        title = "{Validation of the new Hipparcos reduction}",
      journal = {\aap},
         year = 2007,
        month = nov,
       volume = {474},
       number = {2},
        pages = {653-664},
          doi = {10.1051/0004-6361:20078357},
archivePrefix = {arXiv},
       eprint = {0708.1752},
 primaryClass = {astro-ph},
       adsurl = {https://ui.adsabs.harvard.edu/abs/2007A&A...474..653V}
}

@ARTICLE{vida10,
       author = {{Vida}, K. and {Ol{\'a}h}, K. and {K\H{o}v{\'a}ri}, {\protect Zs} and {Jurcsik}, J. and {S{\'o}dor}, {\'A}. and {V{\'a}radi}, M. and {Belucz}, B. and {D{\'e}k{\'a}ny}, I. and {Hurta}, Zs. and {Nagy}, I. and {Posztob{\'a}nyi}, K.},
        title = "{Four-colour photometry of EY Dra: A study of an ultra-fast rotating active dM1-2e star}",
      journal = {Astronomische Nachrichten},
         year = 2010,
        month = mar,
       volume = {331},
       number = {3},
        pages = {250},
          doi = {10.1002/asna.200911341},
archivePrefix = {arXiv},
       eprint = {1001.1852},
 primaryClass = {astro-ph.SR},
       adsurl = {https://ui.adsabs.harvard.edu/abs/2010AN....331..250V}
}

@ARTICLE{vlemmings2024,
       author = {{Vlemmings}, Wouter and {Khouri}, Theo and {Bojnordi Arbab}, Behzad and {De Beck}, Elvire and {Maercker}, Matthias},
        title = "{One month convection timescale on the surface of a giant evolved star}",
      journal = {\nat},
         year = 2024,
        month = sep,
       volume = {633},
       number = {8029},
        pages = {323-326},
          doi = {10.1038/s41586-024-07836-9},
archivePrefix = {arXiv},
       eprint = {2409.06785},
 primaryClass = {astro-ph.SR},
       adsurl = {https://ui.adsabs.harvard.edu/abs/2024Natur.633..323V}
}

@ARTICLE{vogt87,
       author = {{Vogt}, Steven S. and {Penrod}, G. Donald and {Hatzes}, Artie P.},
        title = "{Doppler Images of Rotating Stars Using Maximum Entropy Image Reconstruction}",
      journal = {\apj},
         year = "1987",
        month = "Oct",
       volume = {321},
        pages = {496},
          doi = {10.1086/165647},
       adsurl = {https://ui.adsabs.harvard.edu/abs/1987ApJ...321..496V}
}

@ARTICLE{zahn1977,
       author = {{Zahn}, J. -P.},
        title = "{Tidal friction in close binary systems.}",
      journal = {\aap},
         year = 1977,
        month = may,
       volume = {57},
        pages = {383-394},
       adsurl = {https://ui.adsabs.harvard.edu/abs/1977A&A....57..383Z}
}

@ARTICLE{zhang2000,
       author = {{Zhang}, Zhousheng and {Li}, Yulan and {Tan}, Huisong and {Shan}, Hongguang},
        title = "{BV Photometry of zeta And}",
      journal = {Information Bulletin on Variable Stars},
         year = 2000,
        month = aug,
       volume = {4935},
        pages = {1},
       adsurl = {https://ui.adsabs.harvard.edu/abs/2000IBVS.4935....1Z}
}

@ARTICLE{2012MNRAS.421.1825P,
       author = {{P{\'a}l}, Andr{\'a}s.},
        title = "{FITSH- a software package for image processing}",
      journal = {\mnras},
         year = 2012,
        month = apr,
       volume = {421},
       number = {3},
        pages = {1825-1837},
          doi = {10.1111/j.1365-2966.2011.19813.x},
archivePrefix = {arXiv},
       eprint = {1111.1998},
 primaryClass = {astro-ph.IM},
       adsurl = {https://ui.adsabs.harvard.edu/abs/2012MNRAS.421.1825P}
}
\bibliographystyle{aasjournal}

\end{document}